\documentclass[twocolumn,showpacs,amsmath,amssymb,prb,superscriptaddress]{revtex4-2}

\usepackage{natbib}

\usepackage[version=4]{mhchem}
\usepackage{physics}
\usepackage{siunitx}
\usepackage{graphicx}
\usepackage[table]{xcolor}
\usepackage{lipsum}
\usepackage[T1]{fontenc}
\usepackage[format=hang,justification=centerlast,font={small,stretch=1},labelfont=bf,labelsep=colon]{caption}
\usepackage{subcaption}
\usepackage[backref=true,colorlinks=true,citecolor=blue,urlcolor=blue,linkcolor=blue]{hyperref}

\begin{document}

\title{Tunable magnetocaloric effect and enhanced refrigerant capacity in composite-like oxide heterostructures}
\author{M. Abbasi Eskandari}
\author{S.S. Ghotb}
\affiliation{Institut quantique, Regroupement québécois sur les matériaux de pointe et Département de physique, Université de Sherbrooke, Sherbrooke, J1K 2R1, Québec, Canada}
\author{M. Balli}
\affiliation{LERMA, ECINE, International University of Rabat, Parc Technopolis, Rocade de Rabat-Salé, 11100, Morocco}
\author{P. Fournier}
\affiliation{Institut quantique, Regroupement québécois sur les matériaux de pointe et Département de physique, Université de Sherbrooke, Sherbrooke, J1K 2R1, Québec, Canada}

\begin{abstract}
We report a detailed study of the magnetocaloric effect in heterostructures of \ce{La_{2}NiMnO_{6}} and \ce{La_{2/3}Sr_{1/3}MnO_{3}}. The shape, width and magnitude of the temperature dependence of the magnetic entropy change ($\Delta S_{\mathrm{m}}$) for these multilayer samples can be tuned and controlled by changing their layout. A large $\Delta S_{\mathrm{m}}$ over a wide temperature range which goes beyond room temperature is observed in all samples. We observe a temperature-independent table-top-like $\Delta S_{\mathrm{m}}$ over a temperature range as large as \SI{100}{\kelvin} in the trilayer samples. \ce{La_{2}NiMnO_{6}} double perovskite with multiple magnetic phase transitions is the key to tuning and shaping the temperature dependence of the magnetic entropy changes to suit the requirements of several cooling cycles. 
\end{abstract}
\maketitle

\section{Introduction}

Magnetic refrigeration technology based on the magnetocaloric effect (MCE) has been under the spotlight due to its advantages over the conventional gas compression technology, such as higher energy efficiency and being eco-friendly \cite{balli2017advanced, gschneidnerjr2005recent}. The MCE is an intrinsic property of magnetic materials and it is defined as the heating or cooling of a magnetic substance as it is magnetized or demagnetized, respectively. The MCE was first reported in 1917 by P. Weiss and A. Piccard in nickel close to its paramagnetic to ferromagnetic transition temperature. They noted that \ce{Ni} would warm up or cool down when magnetized or demagnetized \cite{weiss1917phenomene}. Isothermal magnetic entropy change ($\Delta S_{\mathrm{m}}$) and adiabatic temperature change ($\Delta T_{\mathrm{ad}}$) are two parameters that are mostly used in order to characterize the potential of a magnetocaloric material. $\Delta S_{\mathrm{m}}$ and $\Delta T_{\mathrm{ad}}$ indicate the amount of heat that can be moved during the refrigeration process and the temperature change that can be achieved, respectively. They can be measured using specific heat or magnetization. In the late \num{1970}s, gadolinium was introduced as the first working material for magnetic refrigeration taking advantage of its large MCE near its magnetic transition close to room temperature \cite{brown1976magnetic}. \ce{Gd} shows a large isothermal entropy change of $-\Delta S_{\mathrm{m}}=~$\SI{5.5}{\joule\per\kilogram\per\kelvin} and an adiabatic temperature change of $\Delta T_{\mathrm{ad}}=~$\SI{6}{\kelvin} near its transition temperature at \SI{294}{\kelvin} under a magnetic field of \SI{2}{\tesla}, numbers still used for comparison with novel materials. However, for an implementation in domestic cooling devices, one needs to find materials with a magnetic transition leading to a significant magnetic entropy change over a wide temperature range covering roughly from \SIrange[range-units = repeat]{20}{-20}{\celsius}. For some magnetic refrigeration cycles such as active magnetic regenerative (AMR) refrigeration, a so-called \textit{table-top} temperature independence of the magnetic entropy change would be also a great asset \cite{gschneidnerjr2005recent}.

A narrow working temperature range, the metallic behavior and the high cost of Gd restrict its utilization for domestic and large scale applications. Looking for alternative solutions, many families with significant magnetocaloric effect close to room temperature such as \ce{Gd}-based alloys \cite{pecharsky1997giant, xu2015gdxho1}, \ce{LaFe_{13-x}Si_{x}} intermetallic compounds \cite{balli2009effect, fujieda2002large, fujita2001itinerant} and manganites of general formula \ce{R_{1-x}A_{x}MnO_{3}} (where \ce{R} = Lanthanide and \ce{A} = divalent alkaline earth) have received enormous attention due to their advantages such as tunable transition temperature using chemical manipulations and large MCE \cite{tokura1999colossal, phan2007review}. For instance, magnetic and magnetocaloric properties of \ce{La_{1-x}Sr_{x}MnO_{3}} (\ce{LSMO}) manganites have been widely explored. It shows a high transition temperature ($T_{\mathrm{c}}$), as high as \SI{370}{\kelvin}, with a maximum magnetic entropy change of \SI{1.5}{\joule\per\kilogram\per\kelvin} at \SI{1}{\tesla} \cite{mira2002drop}. Despite the narrow operating temperature range ($\Delta T_{\mathrm{ad}}$) and its metallic nature driven by double exchange, a large $\Delta S_{\mathrm{m}}$ makes \ce{La_{1-x}Sr_{x}MnO_{3}} a promising candidate for magnetic cooling systems at room temperature. Similarly, double perovskites with general formula $\ce{A_{2}BB^{\prime}O_{6}}$ (where \ce{A} is a trivalent rare earth or divalent alkaline and \ce{B} and $\ce{B^{\prime}}$ are transition metals) have also attracted interest \cite{balli2014analysis, brahiti2020analysis, kobayashi1998room}. The main advantage of double perovskites over manganites is their higher electrical resistivity \cite{anderson1963theory, kanamori1959superexchange}. \ce{La_{2}NiMnO_{6}} (\ce{LNMO}) as a near room-temperature ferromagnetic semiconductor is attracting because of its unique properties such as the existence of two magnetic phases with different transition temperatures controlled by the level of $\ce{B/B^{\prime}}$ site cationic ordering \cite{singh2010multiferroic, iliev2009growth, bull2003determination}. Cation-ordered LNMO exhibits a maximum magnetic phase transition at \SI{285}{\kelvin}. In this ordered phase, only ferromagnetic \ce{Ni^{2+}-O-Mn^{4+}} bonds driven by superexchange exist. On the other hand, non-optimal growth conditions can lead to a mixture of ferromagnetic (\ce{Ni^{2+}-O-Mn^{4+}}) and antiferromagnetic (\ce{Ni^{2+}-O-Ni^{2+}} and \ce{Mn^{4+}-O-Mn^{4+}}) bonds in a so-called cation-disordered phase. The fully disordered \ce{LNMO} shows a magnetic transition at $T_{\mathrm{c}}\sim$ \SI{150}{\kelvin} \cite{singh2010multiferroic}. This transition temperature gradually increases with the level of cationic ordering as samples oftentimes show two magnetic transitions. This unique feature allows one to control and tune the magnetic and magnetocaloric properties in \ce{LNMO} by changing the ratio of ordering in the system through variation of the growth conditions in hope of tailoring a proper MCE. For instance, Matte \textit{et al}, \cite{matte2018tailoring} investigated the effect of thin film growth conditions on the level of ordering in \ce{LNMO} and consequently on magnetocaloric properties. The authors were able to control and adjust the shape and width of $\Delta S_{\mathrm{m}}$ and get a temperature-independent magnetic entropy change over a wide temperature range, as large as \SI{100}{\kelvin}, by changing the ratio of ordering in \ce{LNMO}, taking advantage of the two magnetic transitions in the same sample and reproducing in a sense what is expected from a composite.

As other compounds, magnetic and magnetocaloric properties of manganites and double perovskites can be tuned in the hope of achieving a large and temperature independent magnetic entropy change. In order to attain a desirable magnetic entropy change close to room temperature several routes have been widely explored such as doping with different elements and changing the growth conditions \cite{brahiti2020analysis, rebello2011large, matte2018tailoring}. Another approach to control and tune the magnetocaloric properties of materials is to combine different materials in a composite structure \cite{zhong2018table, zhang2016excellent, tian2015achieving}. In this case, materials with different transition temperatures are used to obtain a large magnetic entropy change over a wide temperature range \cite{zhang2016excellent, zhong2018table}. In this paper, we choose heterostructures mimicking composites to achieve a suitable magnetocaloric material at room temperature. For this purpose, we take advantage of the potential presence of tunable multiple magnetic transitions in \ce{LNMO} while combining it to \ce{LSMO} showing a transition above room temperature. We grow and study the properties of bilayers and trilayers with the intention of tailoring the magnetic entropy change over a wide temperature range, while taking advantage of strain effects provoked by the lattice mismatches between the substrates and the films.

\section{Experiments and methods}

\begin{figure}
\center
\includegraphics[scale=0.25]{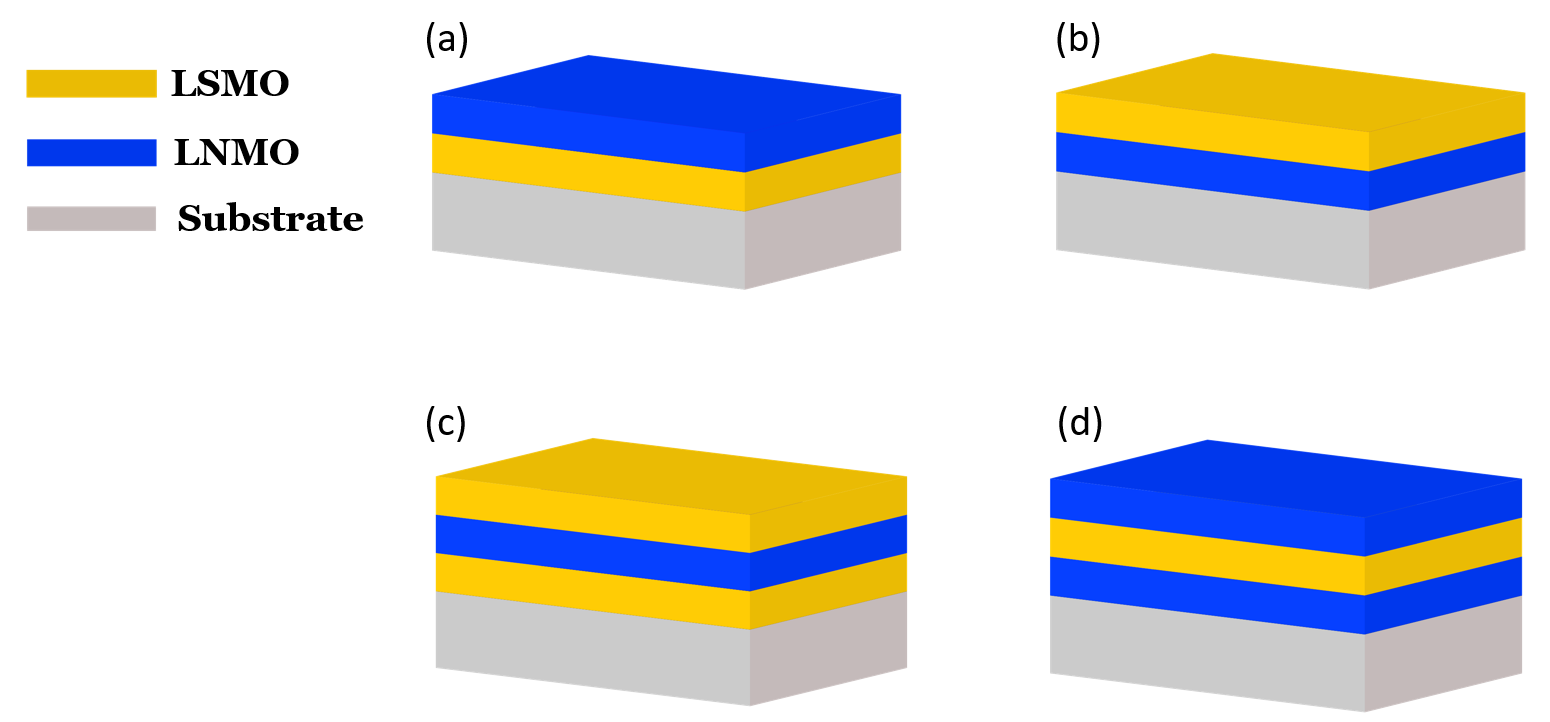}
\caption{Schematic illustration of all grown samples. (a) bilayer of LSMO-LNMO (\ce{B-SN}), (b) bilayer of LNMO-LSMO (\ce{B-NS}), (c) trilayer of LSMO-LNMO-LSMO (\ce{T-SNS}) and (d) trilayer of LNMO-LSMO-LNMO (\ce{T-NSN}).}
\label{Samples}
\end{figure}

In this paper, two series of bilayers and trilayers of \ce{La_{2}NiMnO_{6}} (\ce{LNMO}) and \ce{La_{2/3}Sr_{1/3}MnO_{3}} (\ce{LSMO}) with different configurations are made on (001)-oriented \ce{LSAT} substrates (Figure~\ref{Samples}). The epitaxial bilayers and trilayers samples are grown by pulsed laser deposition (PLD) using a \ce{KrF} excimer laser. The films are deposited at \SI{800}{\celsius}, under an oxygen pressure of \SI{200}{mTorr}. After the deposition, the PLD chamber is filled with oxygen up to \SI{200}{Torr}, and then the films are cooled down to room temperature with a cooling rate of \SI{10}{\celsius/\minute}. In the first type of bilayer samples (Fig.~\ref{Samples}~(a)), a layer of \ce{LSMO} is placed initially on the substrate and then a layer of \ce{LNMO} is deposited on top: it is named \ce{B-SN}. The other bilayer layout is simply the reverse (Fig.~\ref{Samples}~(b)), with a first layer of \ce{LNMO} on the substrate followed by a layer of \ce{LSMO} on top (\ce{B-NS}). These bilayers allow us to sort out the impact of strain on the magnetic properties caused by the lattice mismatch between the substrate and the films. Two different types of trilayer samples were also made. In the first trilayer samples shown in Fig.~\ref{Samples}~(c), a layer of \ce{LNMO} is sandwiched between two \ce{LSMO} layers (\ce{T-SNS}). Fig.~\ref{Samples}~(d) shows the second type of trilayer samples with two \ce{LNMO} layers and a middle \ce{LSMO} layer (\ce{T-NSN}). Each single layer of \ce{LSMO} and \ce{LNMO} has a thickness of \SI{50}{\nm}, implying that the bilayer and trilayer samples have a total thickness of \SIlist[list-units=single]{100;150}{\nm}, respectively. \SI{100}{\nm} thick monolayers were also grown for comparison.

\begin{figure}
\center
\includegraphics[scale=0.4]{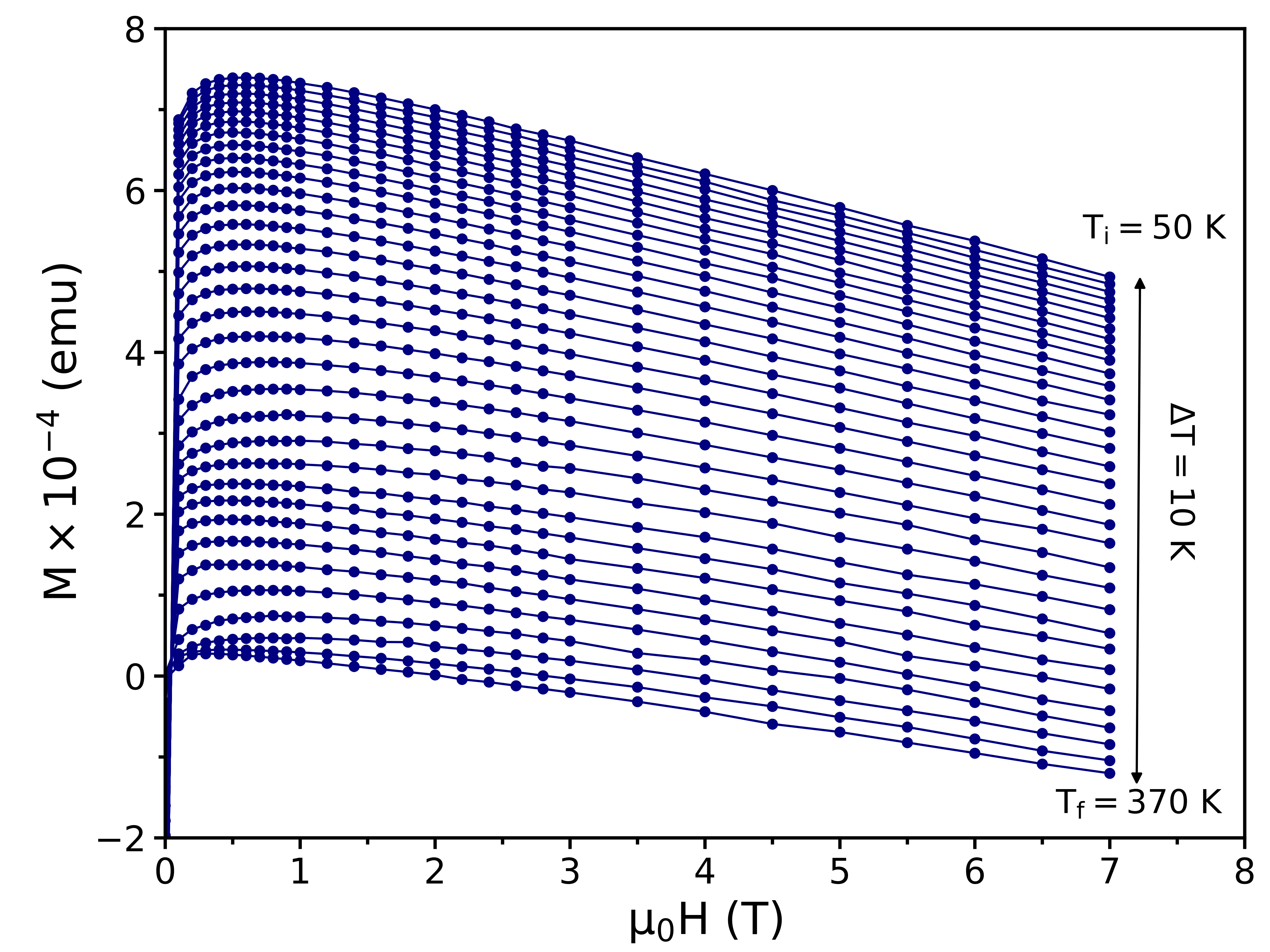}\\
\includegraphics[scale=0.4]{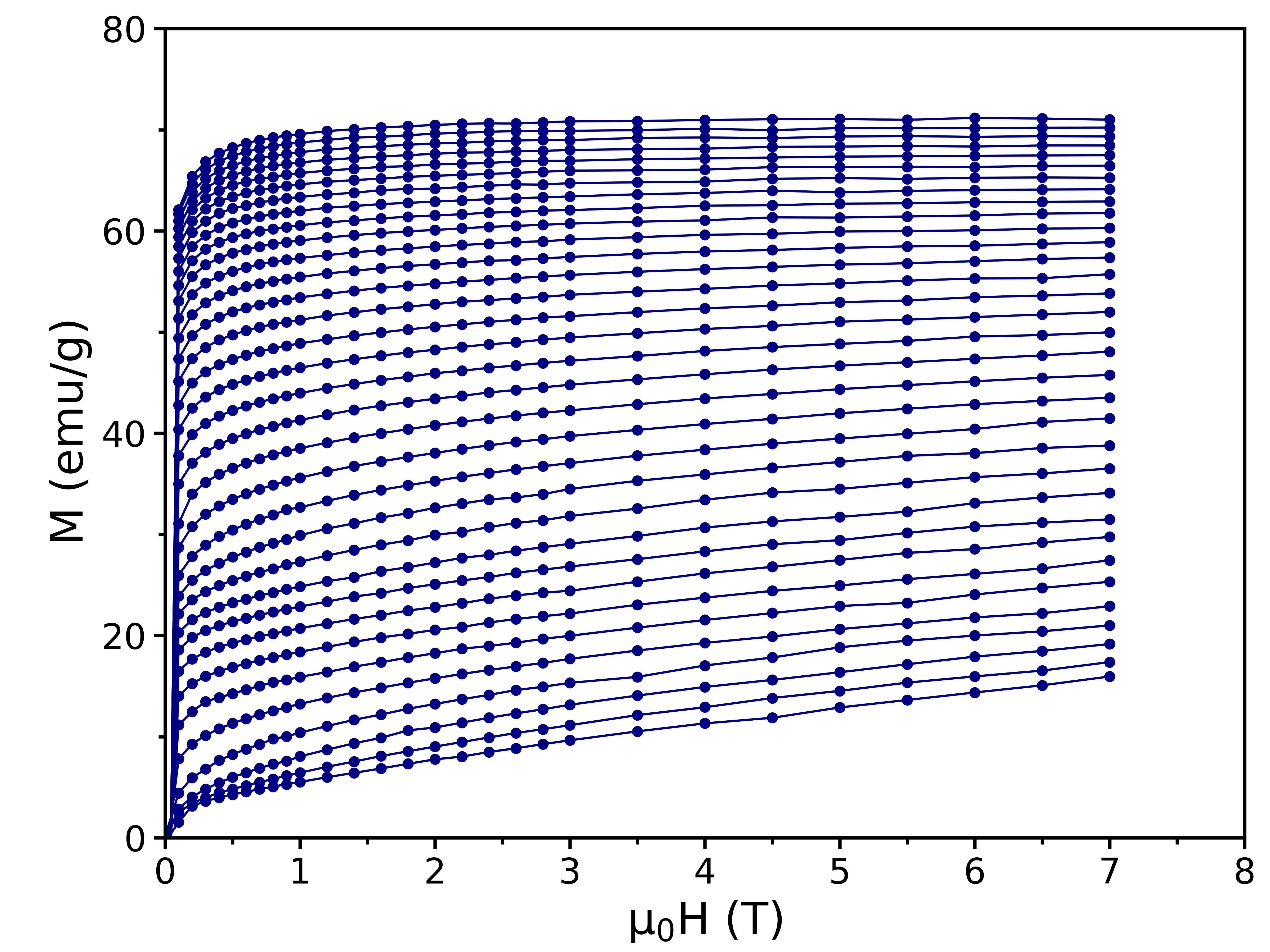}
\caption{Isothermal magnetization curves as a function of magnetic field in the temperature range from \SIrange[range-units=single]{50}{370}{\kelvin} with the temperature interval of \SI{10}{\kelvin} for \ce{B-NS} sample; (a) Including the diamagnetic signal from the substrate and the sample holder (b) After removing the negative background.}
\label{Isotherms}
\end{figure}

Room temperature x-ray diffraction (XRD) was performed using a high-resolution Bruker AXS D8-diffractometer with \ce{CuK_{\alpha 1}} radiation in the $2\theta / \omega$ configuration. The measurement of the magnetic properties are carried out using a Magnetic Property Measurement System (MPMS) from Quantum Design. In order to detect the small magnetic signals from thin films, magnetization measurements were performed using the reciprocating sample option (RSO) with an external magnetic field applied parallel to the surface of the samples. In this work, in order to determine the Arrott plots and the isothermal magnetic entropy changes ($\Delta S_{\mathrm{m}}$), the isothermal magnetization curves as a function of the applied field up to \SI{7}{\tesla} in the temperature range of \SIrange[range-units=single]{50}{370}{\kelvin} with a temperature interval of \SI{10}{\kelvin} are measured for all the samples. Figure~\ref{Isotherms} shows an example of isothermal magnetization measurements for a \ce{B-NS} sample. As one may notice, all the magnetization curves show a negative slope at large field (Fig.~\ref{Isotherms}~(a)) which originates from the diamagnetism of the substrate and the sample holder. As shown in Fig.~\ref{Isotherms}~(b), the magnetization of layers which saturate very rapidly at low field can be clearly seen by removing this negative background. It should also be mentioned that the mass of the multilayers were estimated using the theoretical density of \ce{LSMO} and \ce{LNMO}, as well as the volume of the layers.

%-------------------------XRD-----------------

\section{Results and discussion}

\begin{figure}
\center
\includegraphics[scale=0.4]{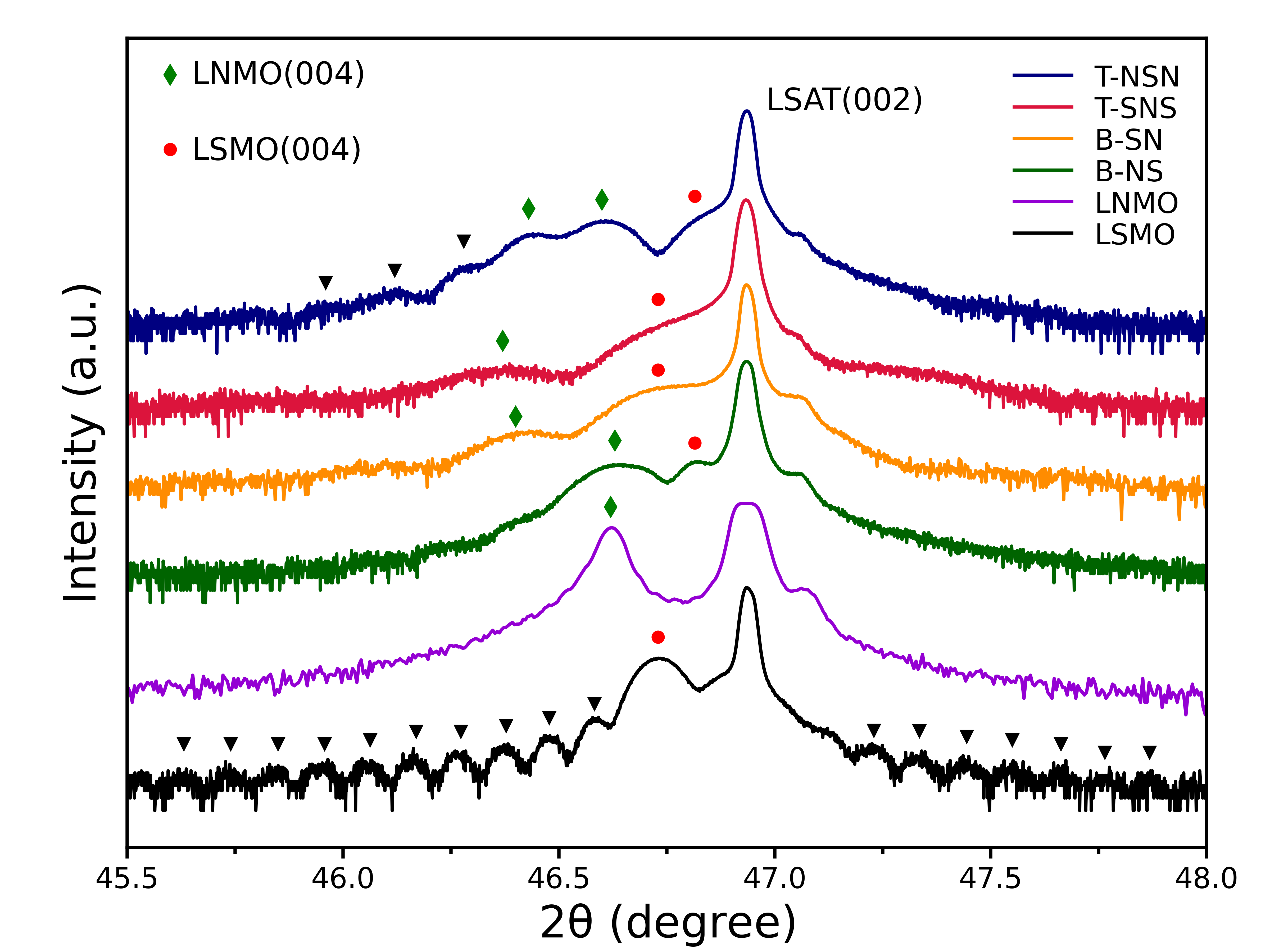}
\caption{X-ray diffraction patterns. From bottom to top: of the monolayers of \ce{LSMO} and \ce{LNMO}, two bilayers of \ce{B-SN} and \ce{B-NS}, as well as two trilayers of \ce{T-SNS} and \ce{T-NSN}. Laue oscillations are specified with $(\blacktriangledown)$. The spectra have been moved vertically for clarity.}
\label{xrd}
\end{figure}

Figure~\ref{xrd} shows the XRD patterns of all samples from \ang{45} to \ang{48} close to the substrate's $(002)$ Bragg peak. The XRD patterns of \SI{100}{\nm} thick monolayers of \ce{LSMO} and \ce{LNMO} at the bottom of the figure show the proximity of their $(004)$ diffraction peaks. In these individual spectra, the single layers of \ce{LSMO} and \ce{LNMO} exhibit a central peak at \SI{46.73}{\degree} and \SI{46.62}{\degree}, respectively. In both cases, their position is in part defined by the impact of strain due to their lattice mismatch with the substrate shifting their angular position with respect to the bulk \cite{guo2013near, ettayfi2016structural}. We are expecting strain to play also a role for thinner layers inserted in multilayers with additional shifts. Nevertheless, the growth conditions used for this work produce extremely smooth \ce{LSMO} as revealed by the numerous Laue oscillations.  

For the multilayers, all observed peaks can be assigned to the $(00l)$ crystallographic planes. It is confirming their out-of-plane orientation and a smooth cube-on-cube growth. For bilayers, the \ce{LNMO} diffraction peak for \ce{B-SN} sits at a lower value of $2\theta$ than in \ce{B-NS} as its top \ce{LNMO} layer is affected by the compressive strain from the bottom \ce{LSMO} layer.  In the supplemental material, we show the spectra close to the substrate's $(001)$ Bragg peak. In this case, we observe Laue oscillations coming from \ce{LSMO} and/or \ce{LNMO} layers. The presence of these oscillations confirms the good crystalline quality of our samples with a well-defined interfaces. The top two spectra in Fig.~\ref{xrd} present the data for \ce{T-NSN} and \ce{T-SNS} trilayers. We notice the presence of three distinct peaks in the spectrum for \ce{T-NSN}, where the peaks for the top and the bottom layers of \ce{LNMO} coexist in the sample: their separation is the result of different strain experienced by both layers. In this particular case, we conclude in fact that the top \ce{LNMO} layer is affected by a stronger compressive strain than the bottom one. Finally, the XRD spectrum for \ce{T-SNS} shows only two diffraction peaks, one is associated to the middle \ce{LNMO} layer, and two \ce{LSMO} layers show a broad peak close to the substrate's peak. 

Atomic force microscopy (AFM) data (see supplemental material) show that the surface roughness on top of the monolayers, bilayers and trilayers is on the order of \SIrange[range-units=single,range-phrase = --]{1}{2}{\nm}, \SIrange[range-units=single,range-phrase = --]{3}{7}{\nm} and \SIrange[range-units=single,range-phrase = --]{7}{14}{\nm}, respectively. In fact, variations of about \SIrange[range-units=single,range-phrase = --]{1}{2}{\nm} over lateral distances of the order of \SI{1}{\mu \m} for the monolayers indicate that the second and/or the third layers grow on top of a fairly smooth surface, resulting in well-defined interfaces between the layers. 

In the next section the magnetic and magnetocaloric properties of bilayer and trilayer samples will be investigated in detail.

\subsection{Bilayers}

\begin{figure}
\center
\includegraphics[scale=0.4]{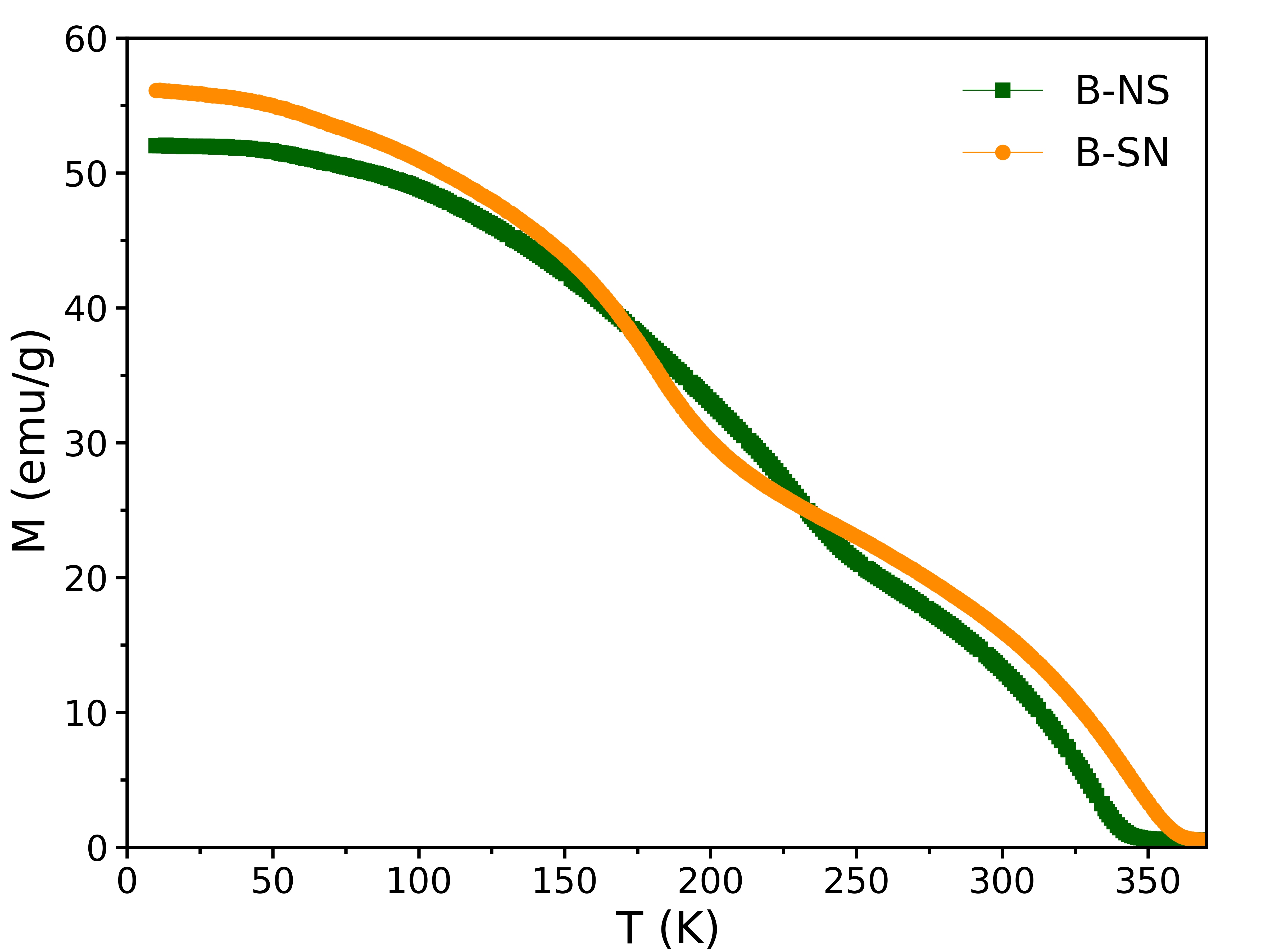}
\caption{Temperature dependence of the FC magnetization of the bilayer samples at a fixed magnetic field of \SI{200}{Oe} in the temperature range between \SIrange[range-units=single]{10}{370}{\kelvin}.}
\label{MvsT_Bilayers}
\end{figure}

Figure~\ref{MvsT_Bilayers} shows field-cooled (FC) magnetization measurements as a function of temperature at a fixed magnetic field of \SI{200}{Oe} for bilayer samples. Both samples undergo two magnetic phase transitions from the ferromagnetic to the paramagnetic state. In the case of the \ce{B-NS} sample, a first transition at roughly \SI{230}{\kelvin} comes from the \ce{LNMO} layer while the other at \SI{330}{\kelvin} is related to the \ce{LSMO} layer. The magnetic transition of \ce{LNMO} at temperatures above \SI{200}{\kelvin} confirms a high but incomplete level of cation ordering at the $\ce{B/B^{\prime}}$ sites in the \ce{LNMO} layer sitting directly on the substrate. For \ce{B-SN} layout, the magnetic transition of the \ce{LSMO} layer shifts at \SI{345}{\kelvin}, while the transition temperature of \ce{LNMO} sitting on top of \ce{LSMO} shifts down to \SI{180}{\kelvin}. The lower transition temperature of the \ce{LNMO} layer in this configuration indicates a lower level of cation ordering than in the previous bilayer \cite{singh2010multiferroic}. This difference in cation ordering level for \ce{LNMO} in both bilayers is likely driven by the different strain fields experienced by the \ce{LNMO} layer \cite{wu2018b, jin2016phase}. Even the slight difference in the transition temperature of the \ce{LSMO} layer in different bilayer configurations with shifts from \SI{330}{\kelvin} in \ce{B-NS} to \SI{345}{\kelvin} in \ce{B-SN} can be also attributed to the effect of strain \cite{dey2007effect, wang2013oxygen}.

This set of bilayers demonstrates that we can change easily the level of cation ordering in \ce{LNMO} and control the magnetic properties of layers just by changing the layout of the layers in bilayers, but also in more complex heterostructures. This feature provides an interesting avenue to tailor the magnetocaloric behavior close to room temperature, which will be discussed in detail later.

\begin{figure}
\center
\includegraphics[scale=0.4]{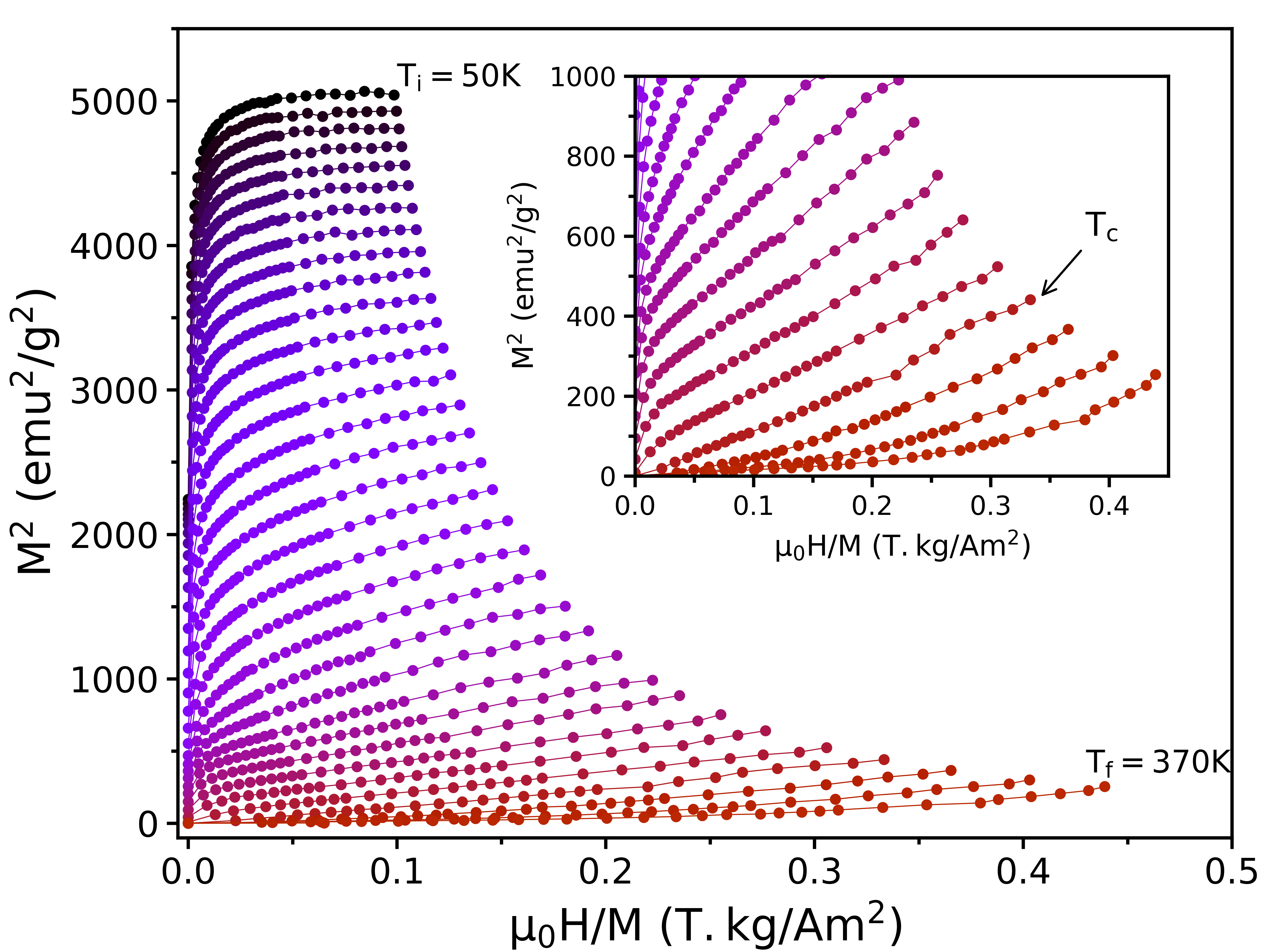}
\caption{Arrott plots in the temperature range from \SIrange[range-units=single]{50}{370}{\kelvin} with a temperature interval of \SI{10}{\kelvin} for \ce{B-NS} sample. The inset of figure shows Arrott plots close to the transition temperature of the sample.}
\label{Bilayers_Arrott_Plots}
\end{figure}

As it is known, MCE strongly depends on the nature of the magnetic phase transition. It should be noted that materials with a first-order magnetic phase transition usually show a large magnetic entropy change in a very narrow temperature range while their magnetic transition is usually accompanied by large thermal and magnetic hysteresis which neither of them are favorable for magnetic cooling systems \cite{phan2007review, smith2012materials}. On the other hand, the broad magnetic phase transition and the reversible nature in second-order phase transition materials make them better suited for magnetic cooling systems. So, in order to get a deeper understanding of the nature of magnetic phase transition in our samples, the Arrott plots for the bilayers are derived from isothermal magnetization curves which are measured in the temperature range between \SIrange[range-units=single]{50}{370}{\kelvin} under to \SIrange[range-units=single,range-phrase = --]{0}{7}{\tesla}. According to Banerjee's criterion \cite{banerjee1964generalised}, the negative or positive slope of Arrott plots indicate whether the magnetic phase transition is first or second order, respectively. The Arrott plots of the \ce{B-NS} bilayer sample is depicted in Figure~\ref{Bilayers_Arrott_Plots} as an example, and it clearly shows a positive slope for the entire temperature range confirming the existence of a second-order magnetic phase transition in our sample. Moreover, the transition temperature of magnetic materials can be extracted using the same plots \cite{arrott1957criterion}. Based on the mean-field theory, Arrott plot curves near the transition temperature become straight lines crossing the origin. The inset of Fig.~\ref{Bilayers_Arrott_Plots} shows the same Arrott plots near the highest transition temperature corresponding to \ce{LSMO} for that \ce{B-NS} sample. Similar plots can be found in the supplemental material for the other configurations. From this data, it was found that the highest transition temperature of \ce{B-NS} and \ce{B-SN} samples occur at \SIlist[list-units=single]{360;340}{\kelvin}, respectively. These transition temperatures from the \ce{LSMO} layers are in close agreement with the M(T) measurements. Unfortunately, the lowest transition temperature of \ce{LNMO} cannot be isolated out of these plots.

In order to determine $-\Delta S_{\mathrm{m}}$, Maxwell's relation linking the magnetic entropy change to the bulk magnetization (M), the temperature (T) and the external magnetic field (B) can be used through the following equation \cite{tishin2016magnetocaloric}:

\begin{equation}
\left(\frac{\partial S}{\partial B}\right)_{P,T} = \left(\frac{\partial M}{\partial T}\right)_{P,B}
\end{equation}

In both isothermal and isobaric conditions, the isothermal magnetic entropy change induced by a change in external magnetic field can then be written as:

\begin{equation}
\Delta S_{\mathrm{m}}(T,B_{f} \rightarrow B_{i}) =  \int_{B_{i}}^{B_{f}} \left(\dfrac{\partial M}{\partial T}\right)_{P,B'} \dd B'
\label{Entropy_int}
\end{equation}

Eq.~\ref{Entropy_int} indicates that the magnetic entropy change is proportional to both the external field variation and the sharpness of a magnetic phase transition. Since magnetization measurements are usually made at discrete field and temperature intervals, $-\Delta S_{\mathrm{m}}$ can rather be computed using \cite{phan2007review}:

\begin{equation}
\Delta S_{\mathrm{m}} (T,B) = \sum_{i} \dfrac{M_{i+1}-M_{i}}{T_{i+1}-T_{i}} \Delta B_{i}
\label{Maxwell}
\end{equation}

where $M_{i}$ and $M_{i+1}$ are the magnetization value measured at temperatures $T_{i}$ and $T_{i+1}$, under a magnetic field changing from \num{0} to $B$, respectively. Thus, Eq.~\ref{Maxwell} (and Eq.~\ref{Entropy_int}) evaluates the surface area between two isothermal magnetization curves measured at $T_{i}$ and $T_{i+1}$ (as presented in Fig.~\ref{Isotherms}~(b)) with $T$ being their average.

The main scope of this work is to tailor a large and temperature independent magnetic entropy change over a wide temperature range. In order to achieve this goal and make a desirable magnetocaloric material, composite-like bilayer and trilayer structures are chosen as the main route. On top of that, it uses \ce{LNMO} as a magnetic material with a tunable magnetic transition and combines it within a composite-like structure in the hope of getting a unique magnetocaloric feature. As discussed earlier, the magnetic properties of bilayer samples such as \(T_{\mathrm{c}}\) are affected by changing their layout. So, in this part we inspect to see how this change affects the magnetocaloric properties of bilayer samples.
 
\begin{figure}
\center
\includegraphics[scale=0.4]{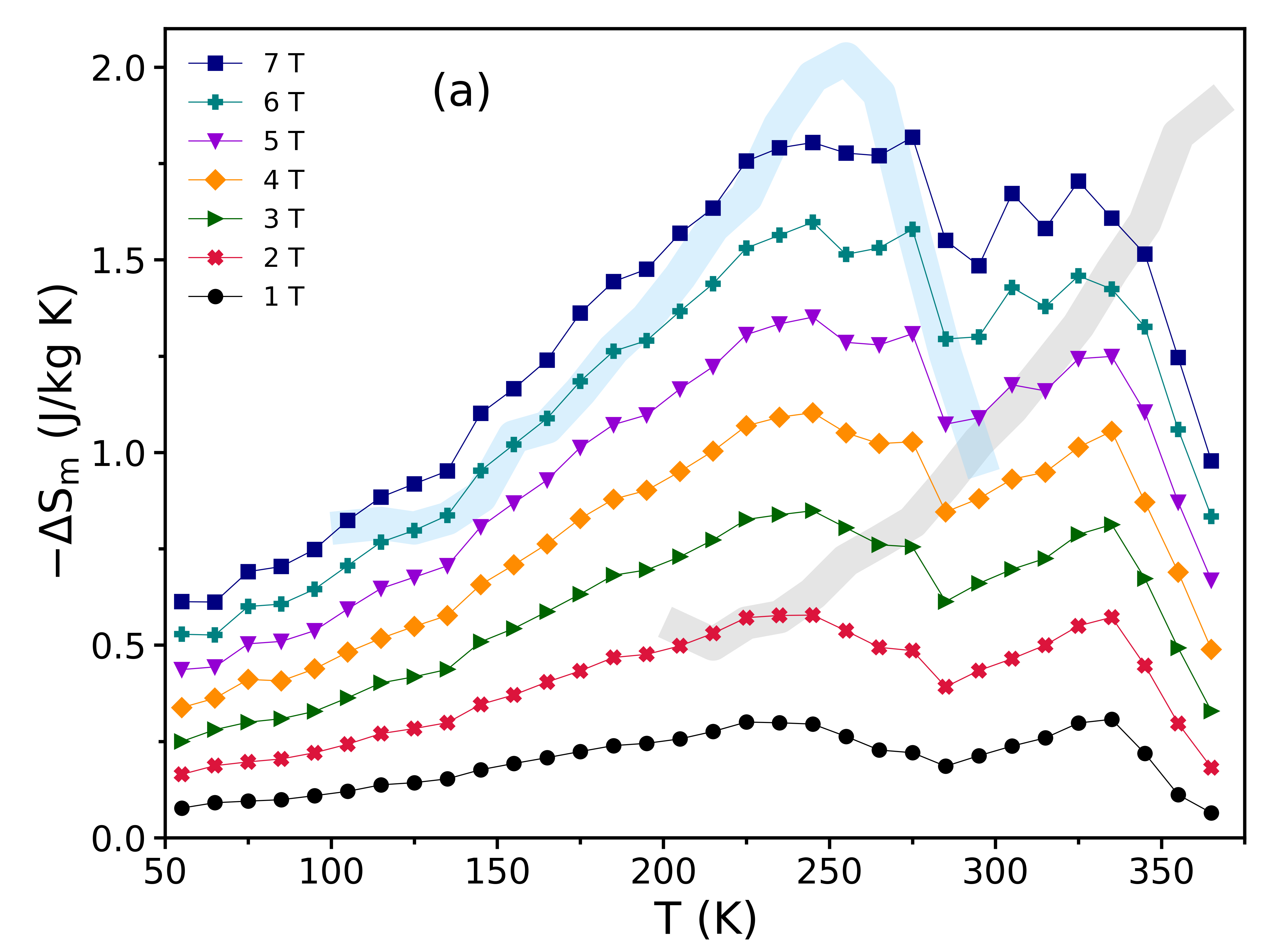}\\
\includegraphics[scale=0.4]{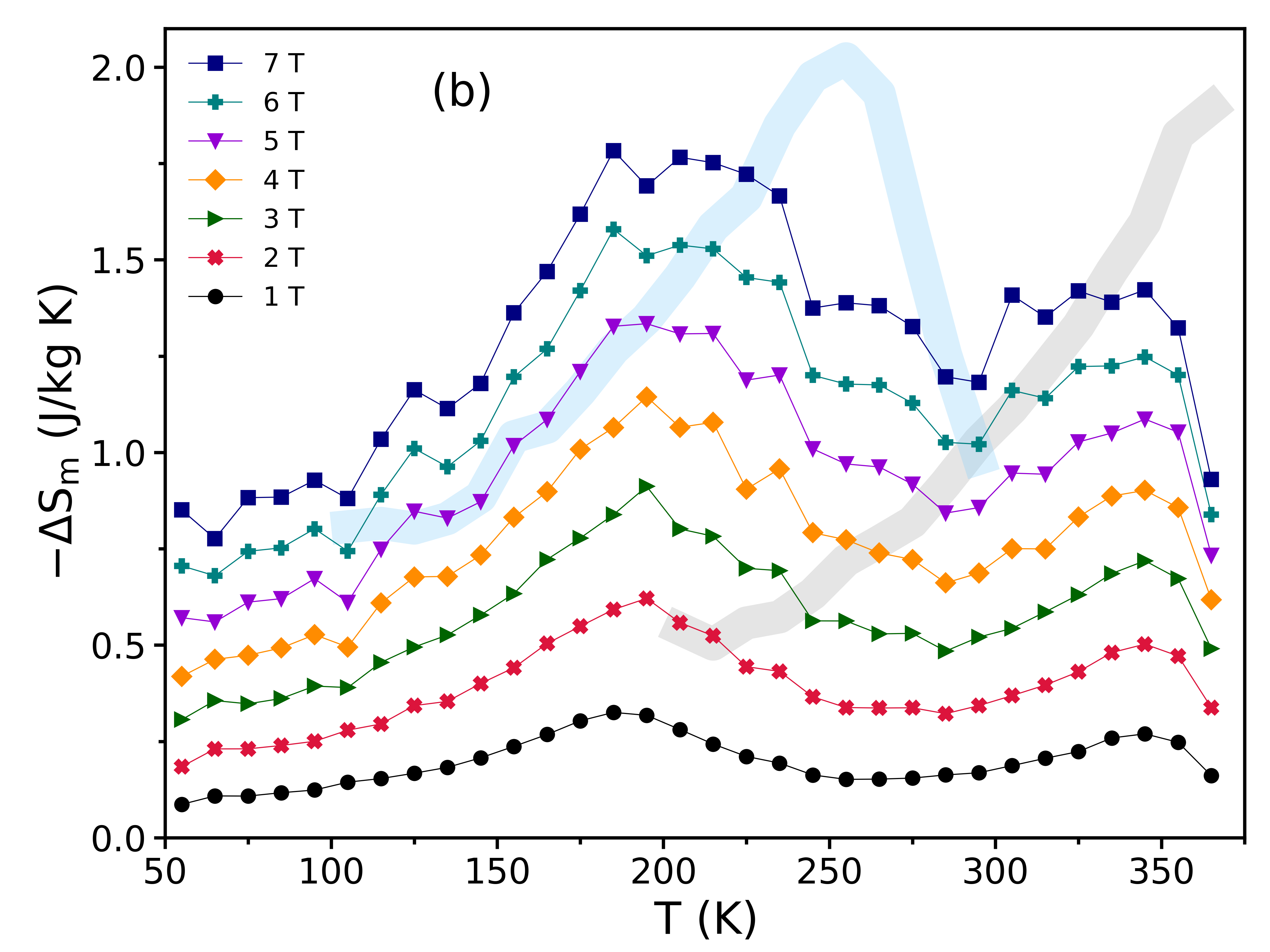}
\caption{The magnetic entropy change as a function of temperature for different magnetic field variations for (a) \ce{B-NS} and (b) \ce{B-SN} samples. The light blue and grey curves on the background show the $-\Delta S_{\mathrm{m}}$ as a function of temperature at a magnetic field of \SI{5}{\tesla} for monolayers of \ce{LNMO} and \ce{LSMO}.}
\label{Bilayers_Entropy}
\end{figure}

The calculated isothermal magnetic entropy changes ($-\Delta S_{\mathrm{m}}$) as a function of temperature are shown in Figures~\ref{Bilayers_Entropy}~(a) and (b) for both bilayer samples under various magnetic field changes up to \SI{7}{\tesla}. Moreover, the magnetic entropy changes of the monolayers of \ce{LNMO} and \ce{LSMO} for $\mu_{0}\Delta H=~$\SI{5}{\tesla} are displayed in the background of Fig.~\ref{Bilayers_Entropy} with light blue and grey, respectively. As expected, both bilayer samples exhibit two broad maxima in $-\Delta S_{\mathrm{m}}$ curves approaching the transition temperatures of each layer. In the \ce{B-NS} sample (Fig.~\ref{Bilayers_Entropy}~(a)), these two maxima seen at \SIlist[list-units=single]{235;335}{\kelvin} are related to the $T_{\mathrm{c}}$ of the cation ordered \ce{LNMO} and the \ce{LSMO} layers, respectively. The transition temperature of the \ce{LNMO} layer almost coincides with that of the monolayer of \ce{LNMO} with a $T_{\mathrm{c}}$ at \SI{245}{\kelvin}, confirming that the level of cationic ordering in this layer is not very much affected by the strain from the top \ce{LSMO} layer. In comparison to the monolayer of \ce{LSMO} in which the $T_{\mathrm{c}}$ is at \SI{365}{\kelvin}, the \ce{LSMO} layer in the \ce{B-NS} sample experiences a \num{30}-\si{\kelvin} shift in transition. The difference could originate from the epitaxial strain caused by the bottom \ce{LNMO} layer leading to a change in the bond length and bond angle of \ce{Mn-O-Mn} bonds in the \ce{MnO_{6}} octahedra. The maximum magnetic entropy changes which are attributed to ordered \ce{LNMO} and \ce{LSMO} layers are \SIlist[list-units=single]{1.80;1.70}{\joule\per\kilogram\per\kelvin} for a magnetic field change of \SIrange[range-units=single,range-phrase=--]{0}{7}{\tesla}. On the other hand, in the \ce{B-SN} sample (Fig.~\ref{Bilayers_Entropy}~(b)), the maximum which is related to the transition of \ce{LNMO} layer shifts down to \SI{185}{\kelvin}, while the one related to the transition of \ce{LSMO} increases to \SI{345}{\kelvin}. The significant drop in the $T_{\mathrm{c}}$ of LNMO can be attributed to the reduction of the cationic ordering level due to the increased epitaxial strain on the \ce{LNMO} layer, which is in agreement with the trend observed in the XRD data (Fig.~\ref{xrd}). Also, $-\Delta S_{\mathrm{m}}$ reaches a maximum value of \SIlist[list-units=single]{1.76;1.42}{\joule\per\kilogram\per\kelvin} for \ce{LNMO} and \ce{LSMO} layers under $\mu_{0}\Delta H = \SI{7}{\tesla}$, respectively.

As noticed, both bilayer samples show a large magnetic entropy change which barely varies over a wide temperature range. The full width at half maximum is about \SI{250}{\kelvin} for both samples (from $\sim$\SI{115}{\kelvin} to above room temperature at $\sim$\SI{365}{\kelvin}). Unlike the other approaches used to tune the magnetic properties and especially the MCE such as doping or changing the growth conditions, this composite-like approach allows one to widen the magnetic entropy change without sacrificing much of the $\Delta S_{\mathrm{m}}$ magnitude. In the next section we show how $\Delta S_{\mathrm{m}}$ can be further improved in terms of intensity and temperature span by adding another layer to the composites structure. Of course, we will show that strain can be used to our advantage as was done with bilayers.

\subsection{Trilayers}

As seen in the previous section, we can take advantage of the sensitivity of cation ordering to strain of \ce{LNMO} in multilayer samples just by changing the stacking order of the layers. This effect originates from different lattice mismatches between the layers and the substrate. It provides an opportunity to control the magnetic and magnetocaloric properties of more complex heterostructures and tailor a large MCE over a wide temperature range. As a proof of concept, trilayer samples with \ce{LNMO-LSMO-LNMO} and \ce{LSMO-LNMO-LSMO} configurations are compared.

\begin{figure}
\center
\includegraphics[scale=0.4]{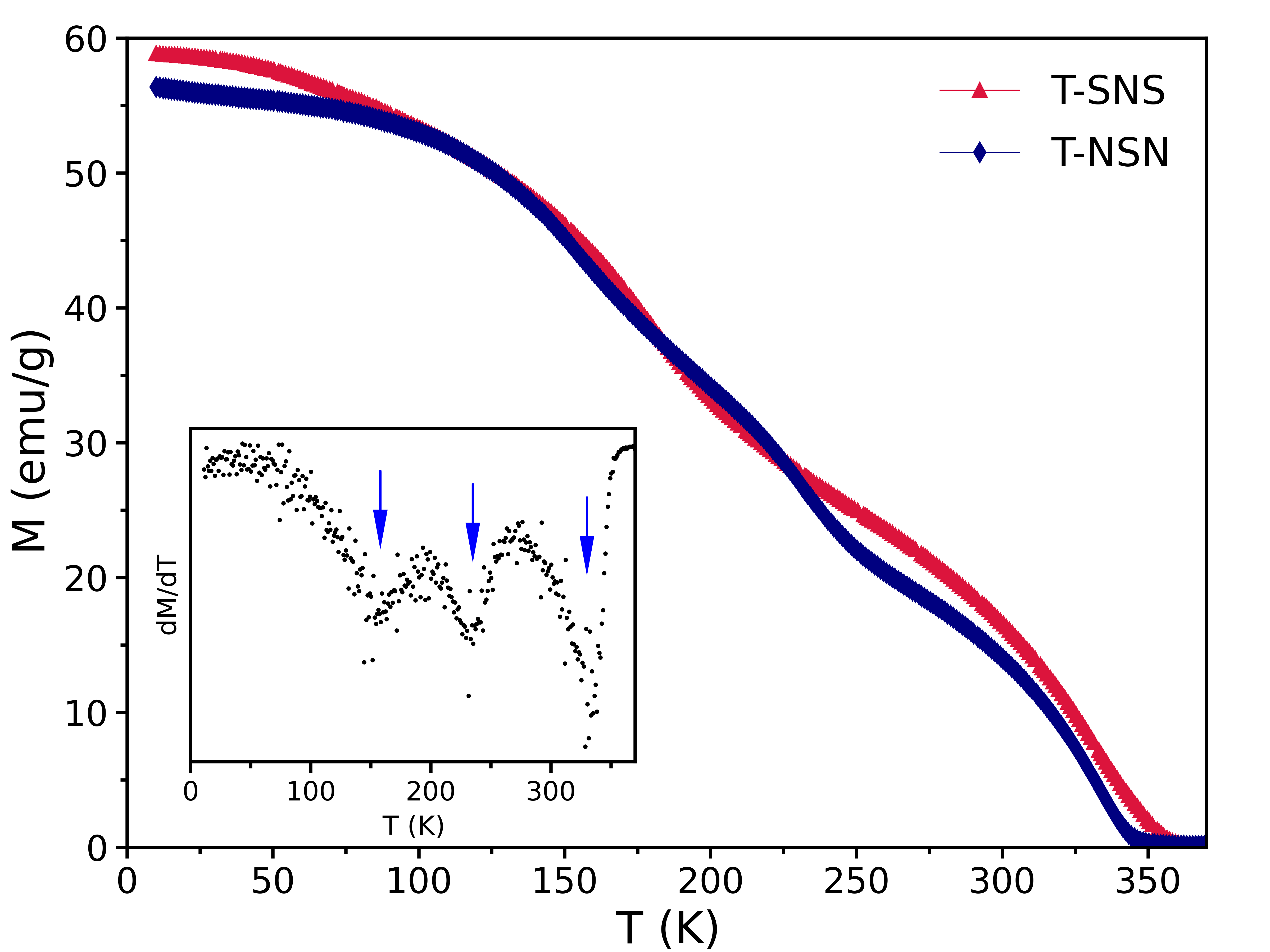}
\caption{Magnetization as a function of temperature under a magnetic field of \SI{200}{Oe} for the trilayer samples. The inset shows the derivative of the magnetization with respect to the temperature for the \ce{T-NSN} sample. The transitions are indicated with blue arrows.}
\label{MvsT_Trilayers}
\end{figure}

Figure~\ref{MvsT_Trilayers} shows field-cooled (FC) magnetization as a function of temperature under a fixed magnetic field of \SI{200}{Oe} for both trilayer samples. As expected, \ce{T-NSN} sample exhibits three magnetic phase transitions at roughly \SIlist[list-units=single]{155;230;335}{\kelvin}. This is confirmed in particular using the derivative $\dd M/\dd T$ as shown in the inset of Fig.~\ref{MvsT_Trilayers}. The first transition at \SI{155}{\kelvin} comes from the top \ce{LNMO} layer experiencing strain similar to that in the \ce{B-SN} sample. The second magnetic transition at \SI{230}{\kelvin} comes from the bottom \ce{LNMO} layer in direct contact with the substrate as was observed in the \ce{B-NS} bilayer. Finally, the transition at \SI{335}{\kelvin} is related to the \ce{LSMO} layer sitting between two \ce{LNMO}, consistent also with its $T_{\mathrm{c}}$ observed in the \ce{B-NS} bilayer. In the other trilayer sample (\ce{T-SNS}), we observe only two magnetic phase transitions in the M(T) curve at \SIlist[list-units=single]{180;340}{\kelvin}. The \ce{LSMO} layers show a $T_{\mathrm{c}}$ at \SI{340}{\kelvin} and the middle \ce{LNMO} layer has a $T_{\mathrm{c}}$ at \SI{180}{\kelvin}. This value of $T_{\mathrm{c}}\sim$\SI{180}{\kelvin} indicates that it has almost the same level of cationic ordering as what we have observed for \ce{LNMO} in \ce{B-SN} sample.

In a trilayer sample with two \ce{LNMO} layers, we can have both disordered and highly ordered \ce{LNMO} phases simultaneously in the sample from the top and bottom layers experiencing different strain fields. As a result, trilayer samples with the \ce{T-NSN} configuration gives three different transitions which covers quite a large temperature range starting from \SIrange[range-units=single]{155}{335}{\kelvin}. With an appropriate choice of thicknesses for each layer, one could get a temperature-independent table-top magnetic entropy change over a wide temperature range close to room temperature.

\begin{figure}
\center
\includegraphics[scale=0.4]{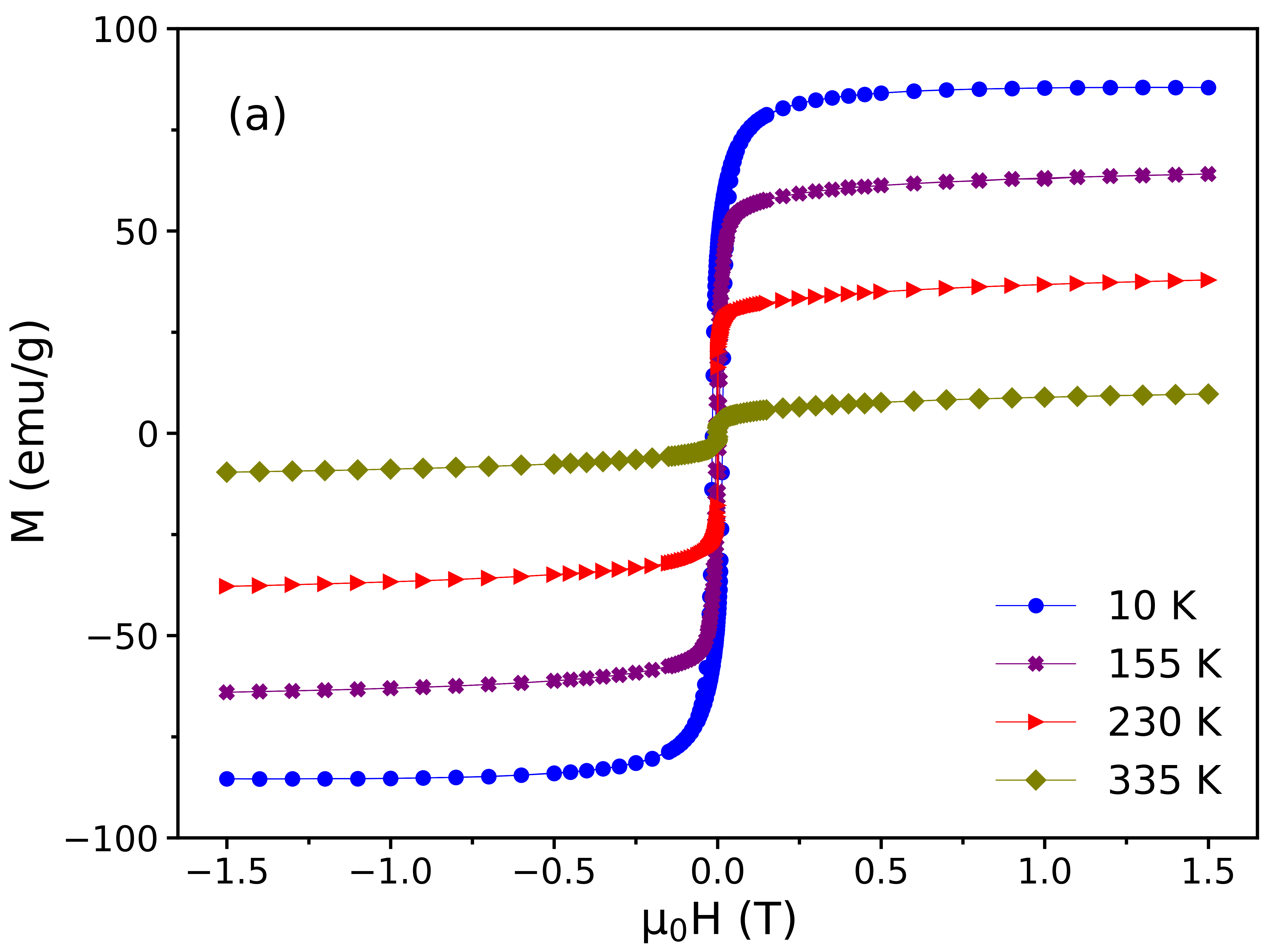}\\
\includegraphics[scale=0.4]{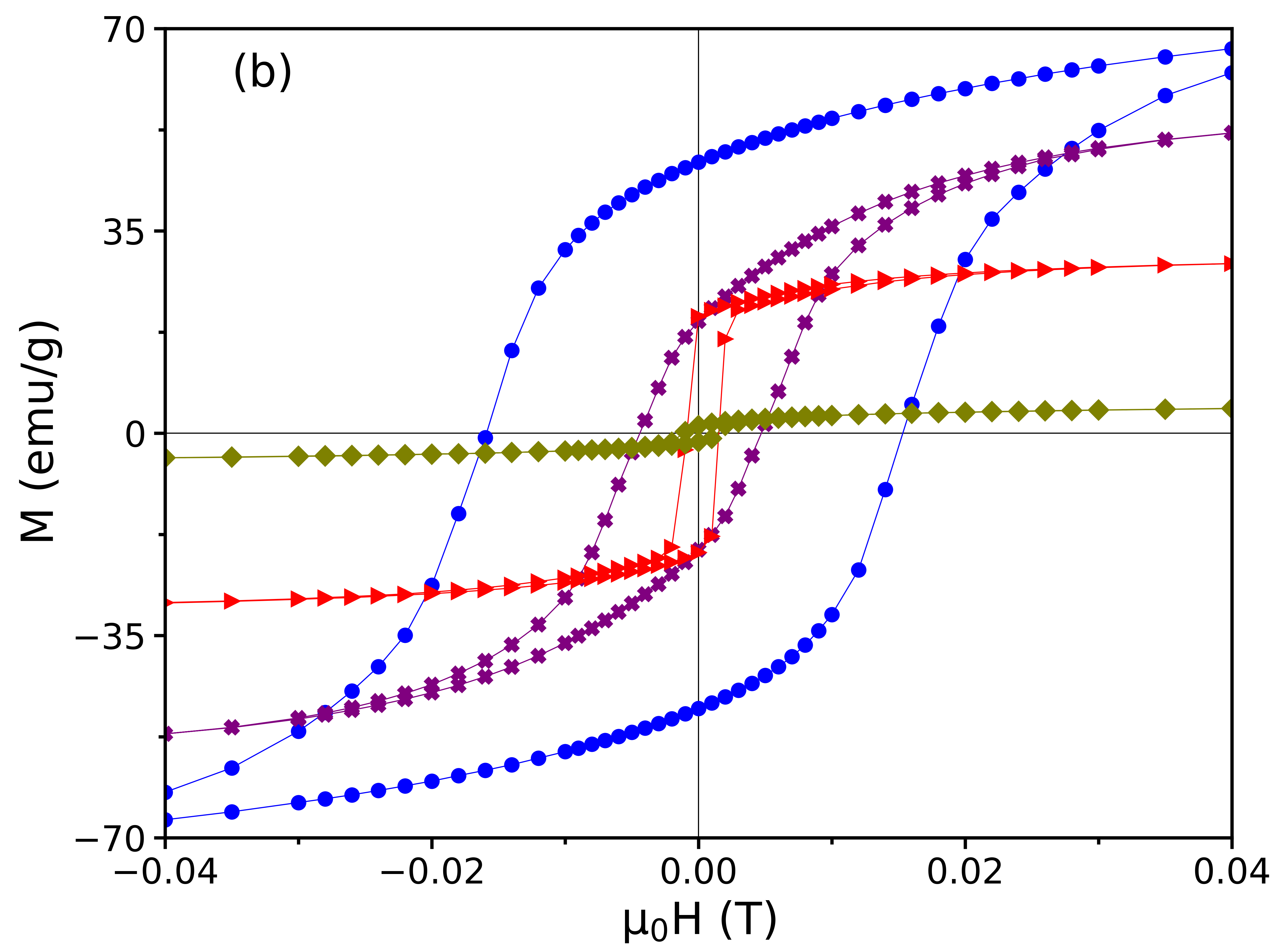}
\caption{(a) Magnetization as a function of magnetic field for a \ce{T-NSN} sample at \SIlist[list-units=single]{10;155;230;335}{\kelvin}. (b) The hysteresis loops at low field showing the small coercive fields.}
\label{MvsH_Hysteresis}
\end{figure}

As mentioned before, one of the biggest advantage of SOMT materials is their small thermal and magnetic hysteresis. In Figure~\ref{MvsH_Hysteresis} (a), the magnetization as a function of magnetic field was measured for a \ce{T-NSN} sample at the transition temperature of each layer and at \SI{10}{\kelvin}. The sample saturates very quickly at low magnetic field, and also it exhibits a very low coercive field of \SI{160}{Oe} at \SI{10}{\kelvin} and this value decreases to roughly \SI{10}{Oe} at \SI{335}{\kelvin} (Fig.~\ref{MvsH_Hysteresis}~(b)). The same is observed for the \ce{T-SNS} sample (see supplemental material).

\begin{figure}
\center
\includegraphics[scale=0.4]{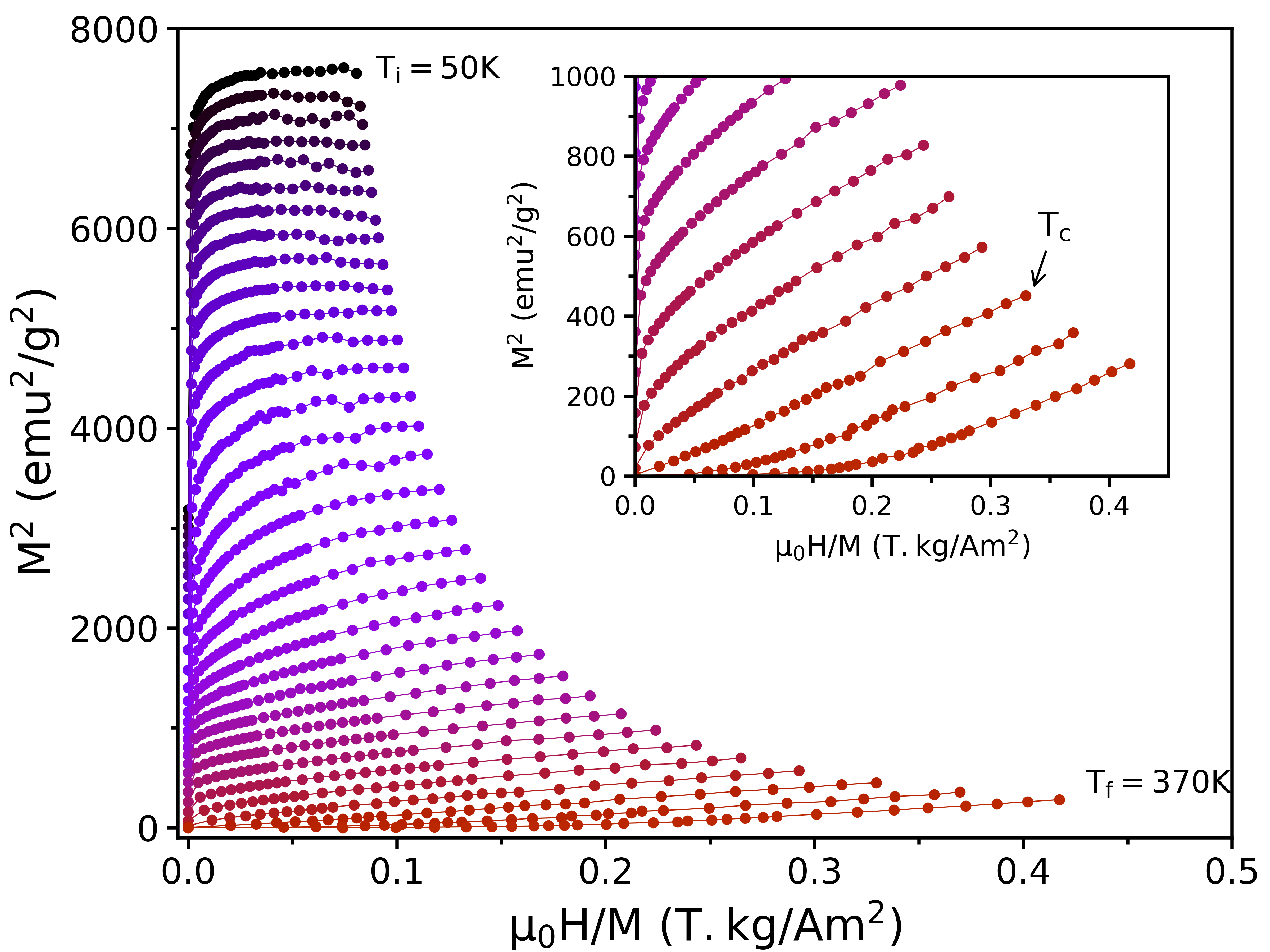}
\caption{Arrott plots of \ce{T-SNS} sample in the temperature range between \SIlist[list-units=single]{50;370}{\kelvin}. The inset shows Arrott plots close to the $T_{\mathrm{c}}$ of the \ce{LSMO} layer.}
\label{Trilayers_Arrott_Plot}
\end{figure}

Magnetic isotherms measurements are carried out in the temperature range going from \SIrange[range-units=single]{50}{370}{\kelvin} to determine Arrott plots and magnetic entropy changes for both trilayer samples. Similar to bilayers, an example of Arrott plots measurements for trilayers is shown in Figure~\ref{Trilayers_Arrott_Plot}. All $M^{2}$ vs. $H/M$ curves of the \ce{T-SNS} sample show positive slopes in the entire temperature range which indicates that the sample undergoes a second-order magnetic phase transition. A similar trend can be observed in the Arrott plots of the other trilayer (\ce{T-NSN}, see supplemental material). Furthermore, in order to determine the $T_{\mathrm{c}}$ of the samples, the Arrott plots close to the transition temperature of \ce{T-SNS} multilayer is depicted in the inset of Fig.~\ref{Trilayers_Arrott_Plot}. It is found that the transition of \ce{T-SNS} and \ce{T-NSN} samples occur at \SIlist[list-units=single]{350;340}{\kelvin}, respectively. As we saw in the bilayer samples, these two transitions are those of the \ce{LSMO} layers in agreement with the M(T) measurements in Fig.~\ref{MvsT_Trilayers}.

\begin{figure}
\center
\includegraphics[scale=0.4]{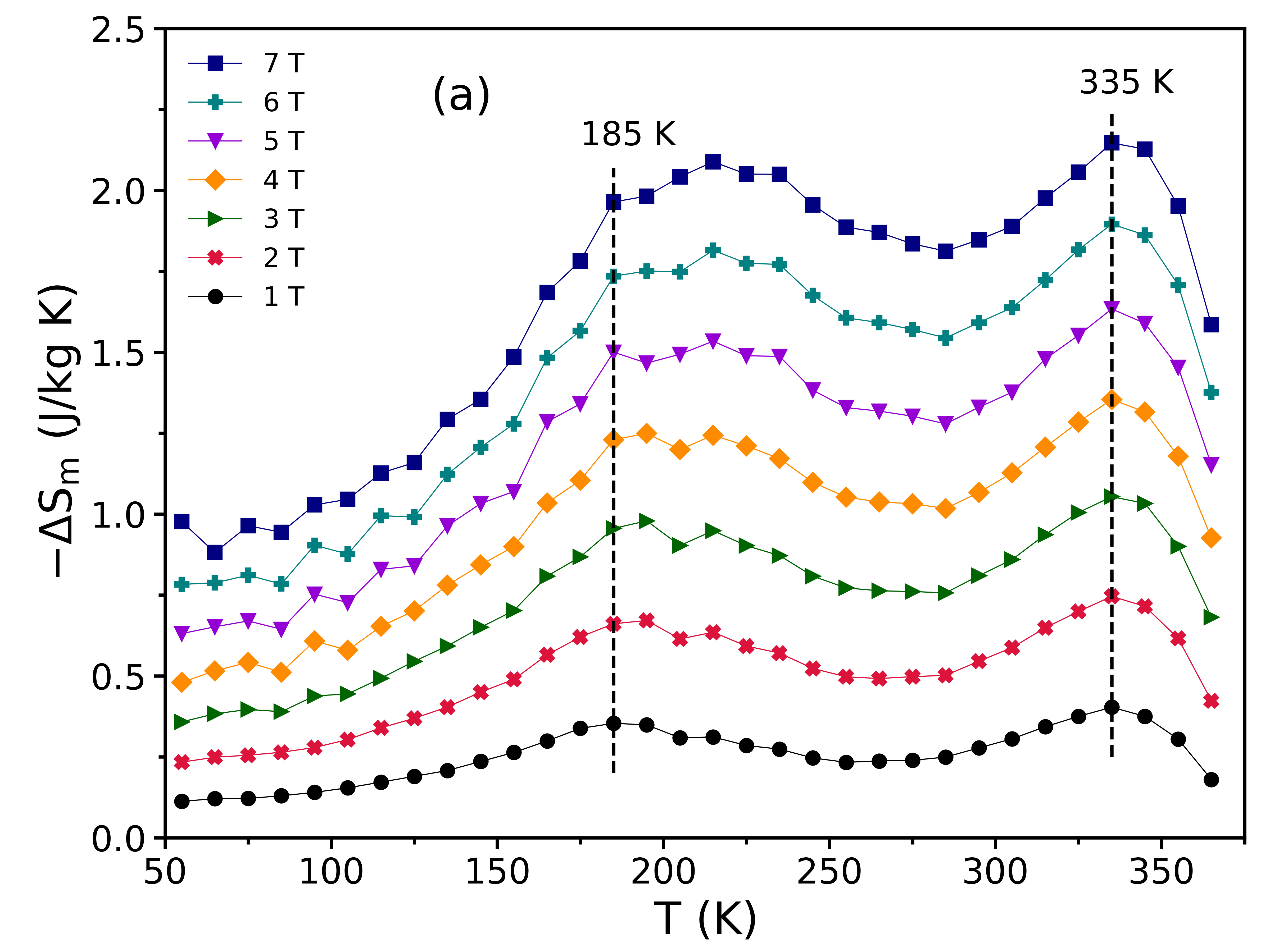}\\
\includegraphics[scale=0.4]{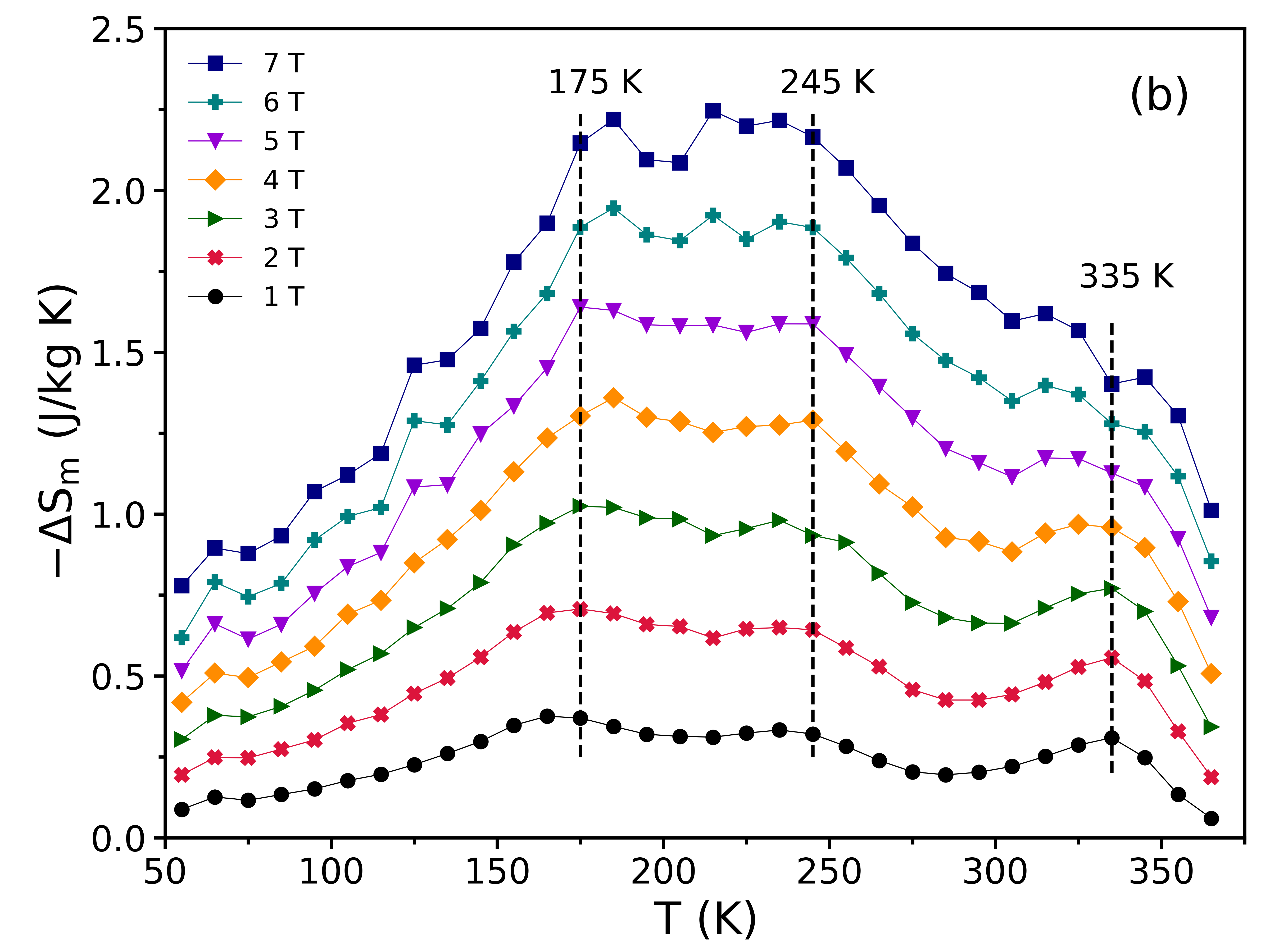}
\caption{Temperature dependence of the magnetic entropy change of (a) \ce{T-SNS} and (b) \ce{T-NSN} samples under various magnetic field changes.}
\label{Trilayers_Entropy}
\end{figure}

Magnetic entropy changes as a function of temperature taken at different magnetic fields for the two trilayer samples are plotted in Figures~\ref{Trilayers_Entropy}~(a) and (b). At the transition temperature of each layer, there is a broad maximum in $\Delta S_{\mathrm{m}}$ that comes from the magnetic phase transitions occurring at the transition temperatures of each layer. For instance, the \ce{T-SNS} sample (Fig.~\ref{Trilayers_Entropy}~(a)) exhibits two peaks in $\Delta S_{\mathrm{m}}$ curves at \SIlist[list-units=single]{185;335}{\kelvin} which correspond to the magnetic transition of \ce{LNMO} and two \ce{LSMO} layers, respectively. The maximum values of $-\Delta S_{\mathrm{m}}$ corresponding to \ce{LNMO} and \ce{LSMO} layers are \SIlist[list-units=single]{1.56;2.03}{\joule\per\kilogram\per\kelvin} for $\mu_{0}\Delta H =~$\SI{7}{\tesla}. As shown in Fig.~\ref{Trilayers_Entropy}~(b), the other trilayer sample (\ce{T-NSN}) shows three peaks in $\Delta S_{\mathrm{m}}$ curves due to the existence of different magnetic phases in two \ce{LNMO} layers. They can be better observed for low applied field (\SIlist[list-units=single]{1;2}{\tesla}). For this layer configuration, the peaks are located at \SIlist[list-units=single]{175;245;345}{\kelvin} which are related to the transition of disordered \ce{LNMO} layer (the top layer), the cation-ordered \ce{LNMO} layer (the bottom layer) and the \ce{LSMO} middle layer, respectively. For $\mu_{0}\Delta H =~$\SI{7}{\tesla}, $-\Delta S_{\mathrm{m}}$ shows a maximum of \SIlist[list-units=single]{2.21;2.21;1.42}{\joule\per\kilogram\per\kelvin} that correspond to disordered \ce{LNMO}, ordered \ce{LNMO} and \ce{LSMO} layers, respectively.

It must be emphasized that the large magnetic entropy changes with a wide operating temperature range ($\delta T_{\mathrm{FWHM}}$ = \SI{260}{\kelvin}) which go slightly above room temperature would make these multilayer composites an interesting candidate for magnetic cooling systems at room temperature. In addition, both trilayer samples show a fairly flat and temperature-independent $\Delta S_{\mathrm{m}}$ within the mid-range temperature which can extend over $\Delta T\sim$\SI{100}{\kelvin} under $\mu_{0}\Delta H =~$\SI{5}{\tesla} (Figs.~\ref{Trilayers_Entropy}). This temperature independent $\Delta S_{\mathrm{m}}$ over a wide temperature window would make these two trilayer samples suitable candidates for AMR refrigeration \cite{franco2018magnetocaloric}. Moreover, with an appropriate choice of thicknesses for each layer (for example a bit thicker for \ce{LSMO}) one could likely generate a constant $-\Delta S_{\mathrm{m}}$ from \SIrange[range-units=single]{175}{335}{\kelvin}.

A useful tool which allows usually to compare the cooling performances of magnetocaloric materials is the relative cooling power $(RCP)$. $RCP$ is the amount of heat transfer between the cold and hot reservoirs in one refrigeration cycle and it is defined as:

\begin{equation}
RCP = -\Delta S_{\mathrm{m}}(\mathrm{max}) \times \delta T_{\mathrm{FWHM}}
\end{equation}

RCP is not only considering the amplitude of the magnetic entropy change $\Delta S_{\mathrm{m}} (\mathrm{max})$, but it also takes into account the working temperature range $\delta T_{\mathrm{FWHM}}$ which is a key parameter in magnetic cooling systems. $RCP$ is mostly used for materials with a single transition which have a fairly symmetric peak in $\Delta S_{\mathrm{m}} (T)$ curves. However, in the case of composites and materials with more than one transition, refrigerant capacity $(\mathcal{RC})$ is usually employed \cite{matte2018tailoring, zhou2018table}. $\mathcal{RC}$ is specified as the area under the curve of $\Delta S_{\mathrm{m}}$ between the temperatures which correspond to half maximum \cite{gschneidner1999recent}:

\begin{equation}
\mathcal{RC} = -\int_{T_{C}}^{T_{H}} \Delta S_{\mathrm{m}}(T) \dd T
\end{equation}

\begin{figure}
\center
\includegraphics[scale=0.4]{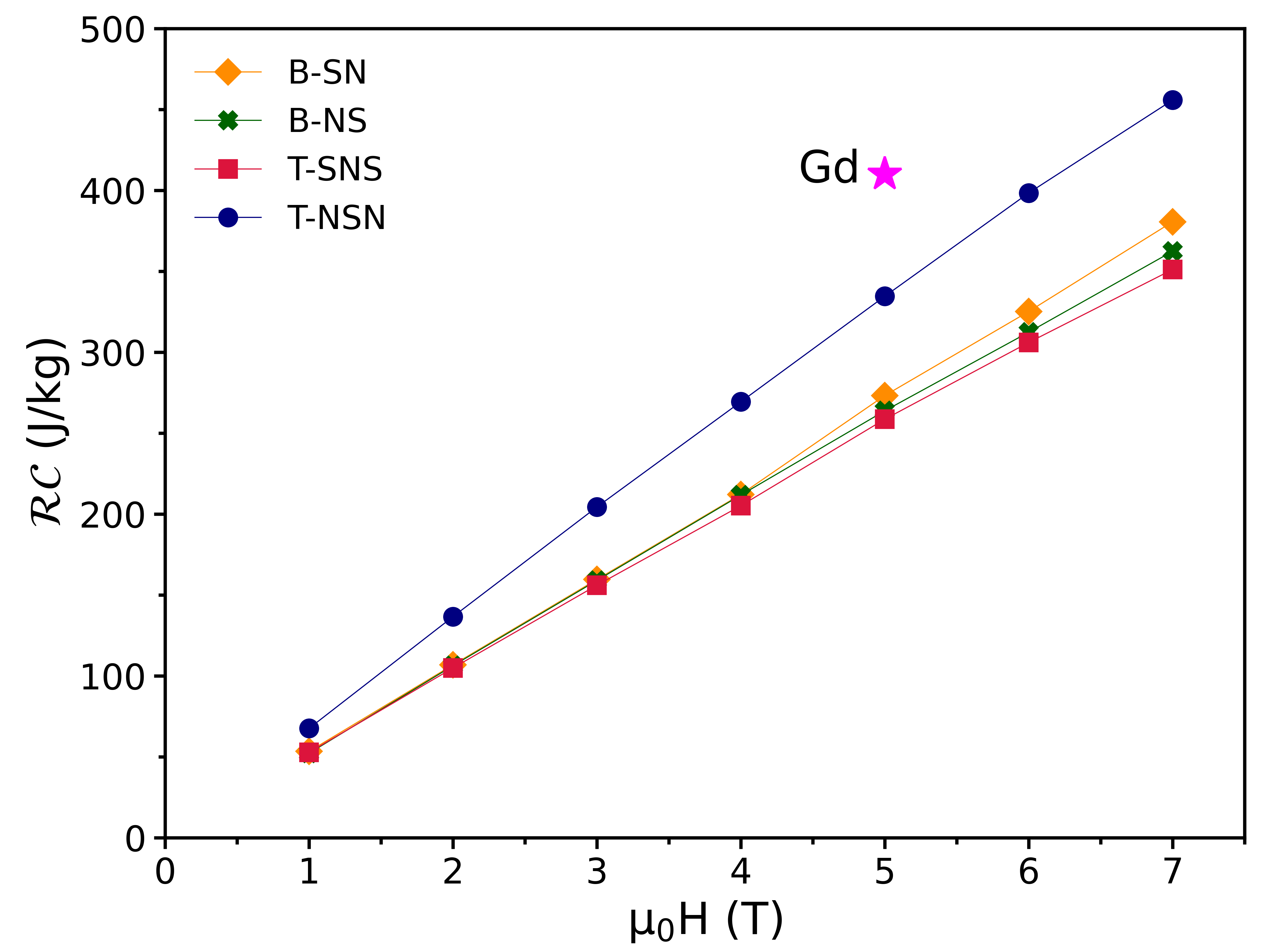}
\caption{Refrigerant capacity ($\mathcal{RC}$) as a function of magnetic field for all samples. $\mathcal{RC}$ of gadolinium is shown for comparison \cite{dan1998magnetic}.}
\label{RC}
\end{figure}

Figure~\ref{RC} presents the $\mathcal{RC}$ values as a function of magnetic field for all four samples and also that of gadolinium as a reference to compare with our samples. All samples show a relatively large $\mathcal{RC}$ with an almost linear magnetic field dependence. The values of $\mathcal{RC}$ reach \SIlist[list-units=single]{273;264;259;335}{\joule\per\kilogram} for \ce{B-SN}, \ce{B-NS}, \ce{T-SNS} and \ce{T-NSN} under a magnetic field of \SI{5}{\tesla}, respectively. By comparing with Gadolinium as a reference \cite{dan1998magnetic}, it can be seen that the value of $\mathcal{RC}$ for \ce{T-NSN} sample is about \SI{82}{\percent} of \ce{Gd} for $\mu_{0}\Delta H =~$\SI{5}{\tesla}. All the values of $\Delta S_{\mathrm{m}}$ and $\mathcal{RC}$ for our samples under different magnetic fields are listed in Table~\ref{All_the_valuse}. Among other families with comparable $\mathcal{RC}$ to \ce{Gd}, such as \ce{La(FeSi)_{13}-based} \cite{shamba2013enhancement, shamba2011reduction} and \ce{Gd_{5}(SiGe)_{4}-based} \cite{provenzano2004reduction} compounds, our multilayer samples have additional advantages over other candidates such as wide working temperature range and a second order magnetic phase transition. Our results confirm that the composite approach combining $3d$ metals oxides such as manganites and double perovskites is a promising path to produce performant cooling devices based on the magnetocaloric effect.

\begin{equation}
\mathcal{RC} = -\int_{T_{C}}^{T_{H}} \Delta S_{\mathrm{m}}(T) \dd T
\end{equation}

\begin{table*}[t]
\center
\setlength{\tabcolsep}{0.4cm}
\renewcommand{\arraystretch}{1.5}
\begin{tabular}{lccccccc}
\hline\hline
&\multicolumn{3}{c}{$-\Delta S_{\mathrm{m}}$ (\si{\joule\per\kilogram\per\kelvin})}&&\multicolumn{3}{c}{$\mathcal{RC}$ (\si{\joule\per\kilogram})}\\
\cline{2-4}\cline{6-8}
Sample &\SI{2}{\tesla}&\SI{5}{\tesla}&\SI{7}{\tesla}&&\SI{2}{\tesla}&\SI{5}{\tesla}&\SI{7}{\tesla}\\
\hline
\ce{B-SN}&0.62&1.33&1.76&&106.82&273.23&380.62\\
\hline
\ce{B-NS}&0.57&1.35&1.80&&106.31&263.81&362.39\\
\hline
\ce{T-SNS}&0.68&1.50&2.03&&104.91&258.61&351.14\\
\hline
\ce{T-NSN}&0.70&1.63&2.21&&136.61&334.57&455.82\\
\hline\hline
\end{tabular}
\caption{The values of $-\Delta S_{\mathrm{m}}$ and $\mathcal{RC}$ for all four multilayers under different magnetic fields.}
\label{All_the_valuse}
\end{table*}

\section{Conclusion}

In this work, we obtain a large and almost temperature independent magnetocaloric effect extending from \SI{175}{\kelvin} up to room temperature in multilayer composites of \ce{LNMO} and \ce{LSMO}. We take advantage of the sensitivity of cationic ordering to strain in \ce{LNMO} layers to achieve multilayers with multiple magnetic transitions. For this purpose, two series of bilayer samples with different layouts of \ce{LSMO-LNMO} and \ce{LNMO-LSMO}, and two series of trilayer samples with layouts of \ce{LSMO-LNMO-LSMO} and \ce{LNMO-LSMO-LNMO} are prepared. All the samples show a large $\Delta S_{\mathrm{m}}$ over a wide temperature range, as large as \SI{260}{\kelvin}, which extends above room temperature. The maximum value of $-\Delta S_{\mathrm{m}}$ in the trilayer samples is found to be \SIlist[list-units=single]{2.03;2.21}{\joule\per\kilogram\per\kelvin} under a magnetic field change of \SI{7}{\tesla} in \ce{T-SNS} and \ce{T-NSN} samples, respectively. Moreover, the trilayer samples reveal an almost temperature independent magnetic entropy change that can extend over a large temperature range of $\Delta T \cong~$\SI{100}{\kelvin}. The refrigerant capacity $(\mathcal{RC})$ reaches a maximum value of \SIlist[list-units=single]{259;335}{\joule\per\kilogram} for $\mu_{0}\Delta H =~$\SI{5}{\tesla} in \ce{T-SNS} and \ce{T-NSN} samples, respectively. These large values of $\mathcal{RC}$ in our composite-like multilayers are comparable to \ce{Gd} and other reference materials. Our results demonstrate that a composite route to design new materials for magnetic cooling systems is a promising path.

\section*{Acknowledgment}

The authors thank B. Rivard, S. Pelletier and M. Dion for technical support. The authors would also like to thank P. Brojabasi for preparing the initial samples. This work is supported by the Natural Sciences and Engineering Research Council of Canada (NSERC) under grant RGPIN-2018-06656, the Canada First Research Excellence Fund (CFREF), the Fonds de Recherche du Québec - Nature et Technologies (FRQNT) and the Université de Sherbrooke. M.B. acknowledges the financial support from the International University of Rabat.

\bibliographystyle{nature}
\bibliography{M.Abbasi_MCE.bib}

%\newpage

	\begin{figure*}
    	
    	\center
	{\Huge Supplemental material}
	
	\end{figure*}

\begin{figure*}
\center
\includegraphics[scale=0.6]{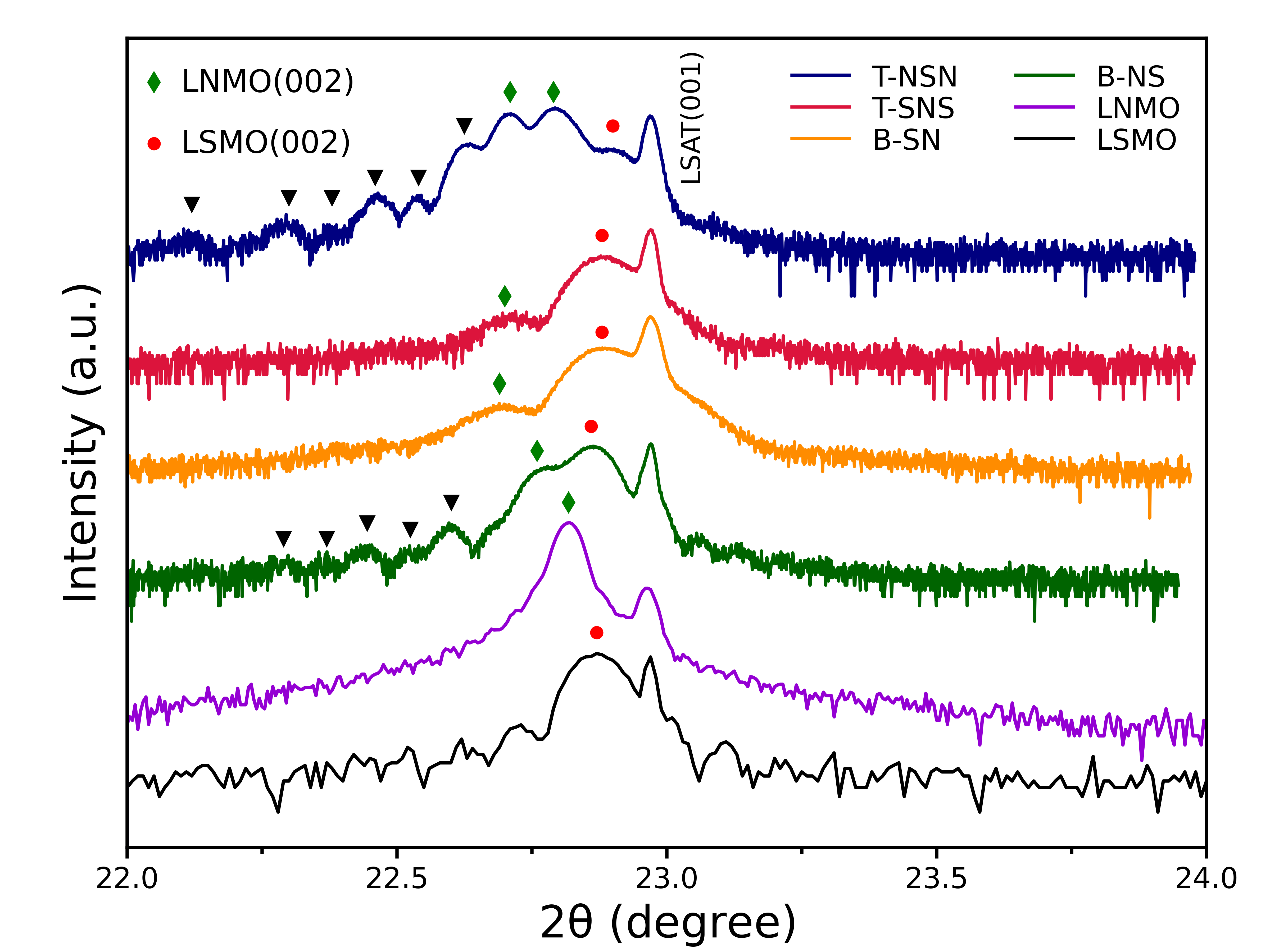}
\caption{X-ray diffraction patterns around the $(001)$ Bragg peak of the substrate from $2\theta =~$\SIrange{22}{24}{\degree}. From bottom to top: the monolayers of \ce{LSMO} and \ce{LNMO}, bilayers of \ce{B-NS} and \ce{B-SN}, and trilayers of \ce{T-SNS} and \ce{T-NSN}. Laue oscillations are specified with $(\blacktriangledown)$. Diamonds indicate the diffraction peaks of \ce{LNMO} and solid circles for \ce{LSMO}.}
\end{figure*}

\begin{figure*}

	\center
	\begin{subfigure}{\textwidth}
	\center
	\includegraphics[width=\textwidth]{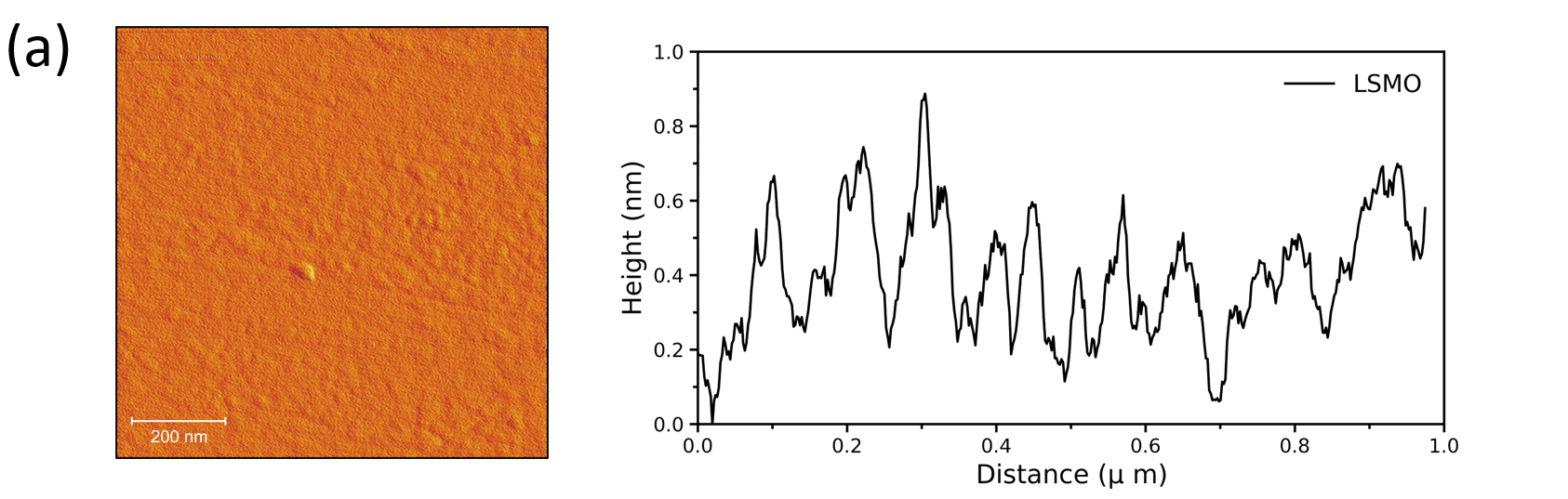}
	\end{subfigure}
\end{figure*}
\begin{figure*}
\ContinuedFloat
	\begin{subfigure}{\textwidth}
	\center
	\includegraphics[width=\textwidth]{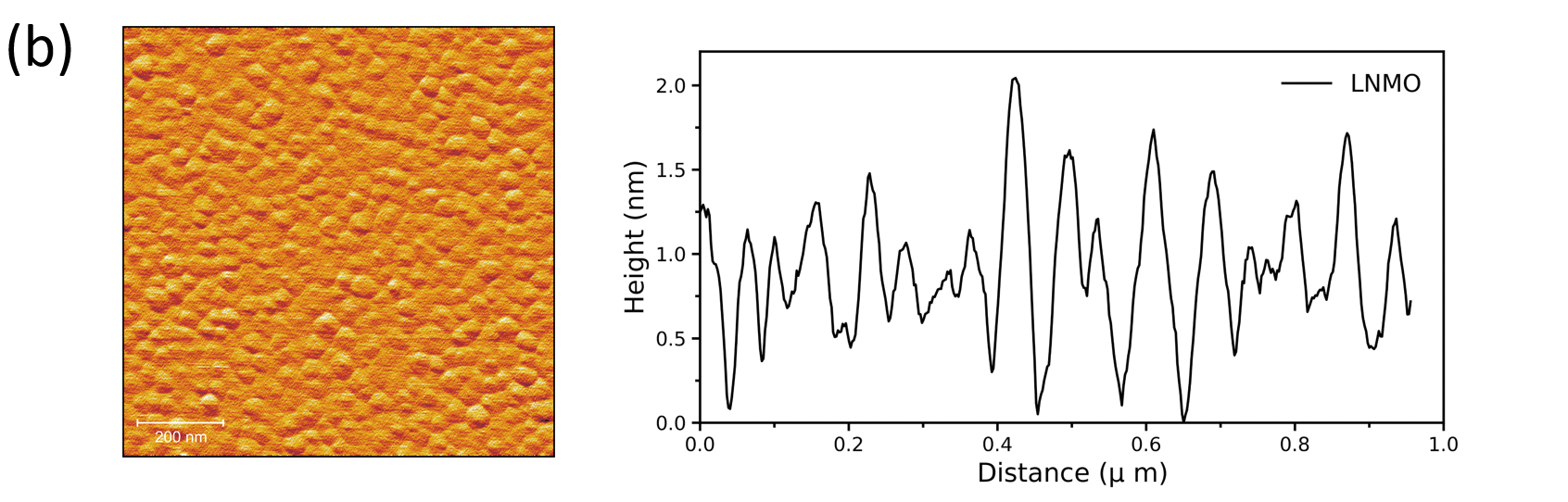}
	\end{subfigure}
\end{figure*}
\begin{figure*}
\ContinuedFloat
	\begin{subfigure}{\textwidth}
	\center
	\includegraphics[width=\textwidth]{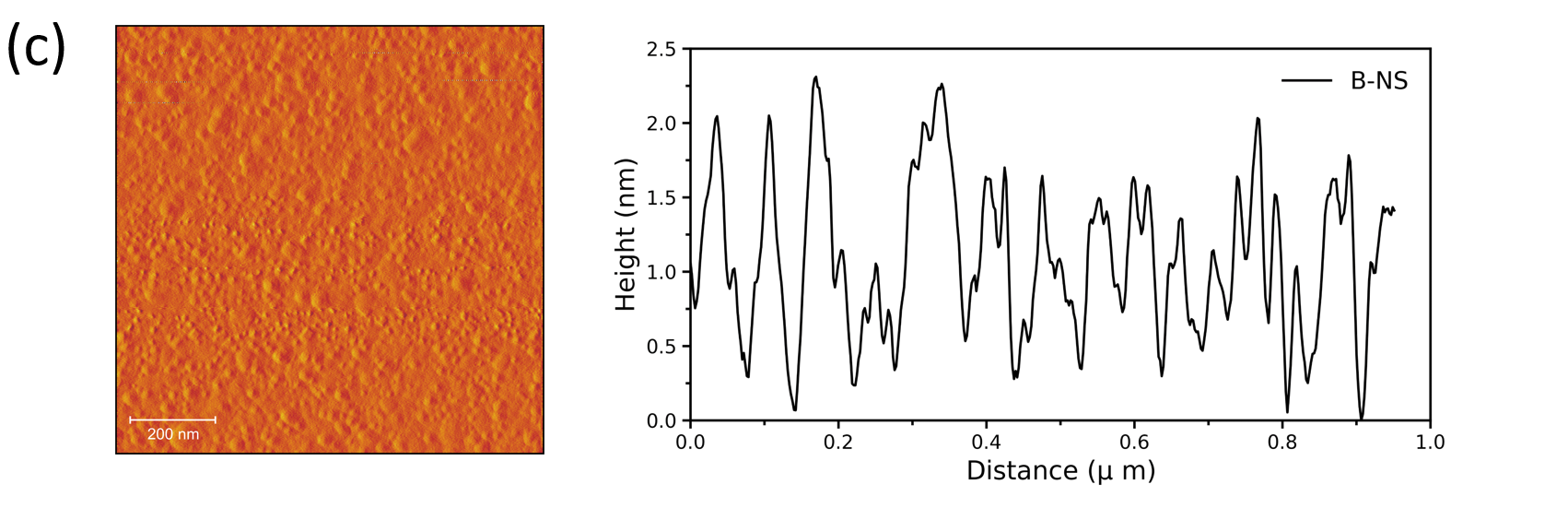}
	\end{subfigure}
\end{figure*}

\begin{figure*}
\ContinuedFloat
	\begin{subfigure}{\textwidth}
	\center
	\includegraphics[width=\textwidth]{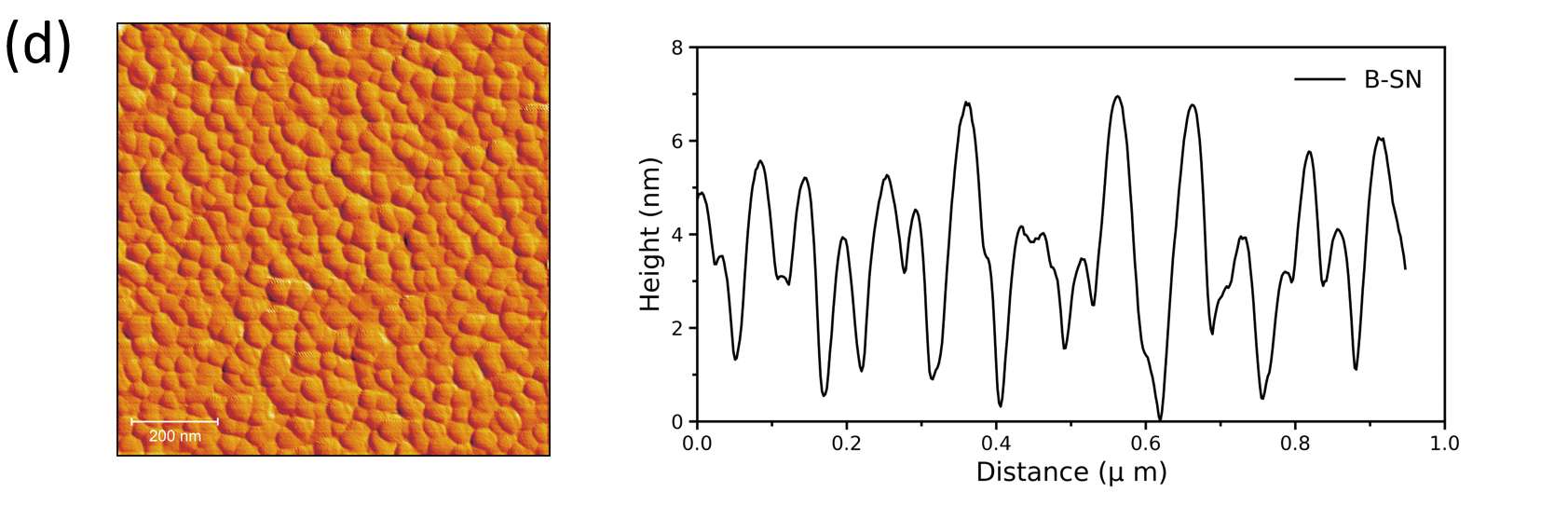}
	\end{subfigure}
\end{figure*}
	
\begin{figure*}
\ContinuedFloat
	\begin{subfigure}{\textwidth}
	\center
	\includegraphics[width=\textwidth]{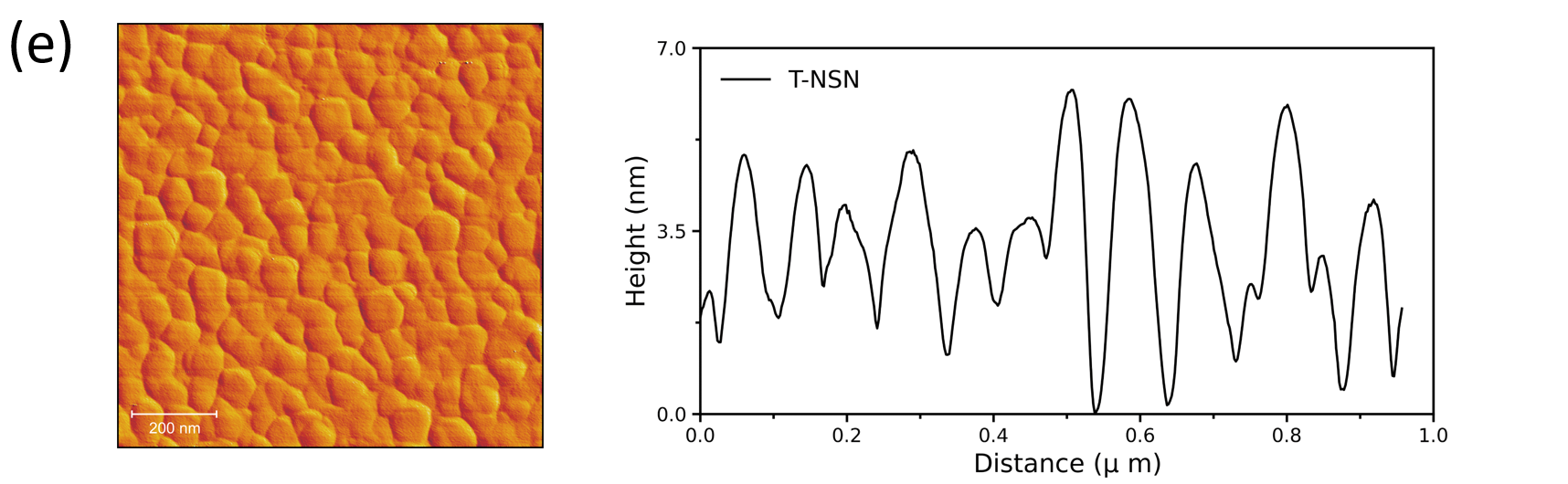}
	\end{subfigure}
\end{figure*}

\begin{figure*}
\ContinuedFloat
	\begin{subfigure}{\textwidth}
	\center
	\includegraphics[width=\textwidth]{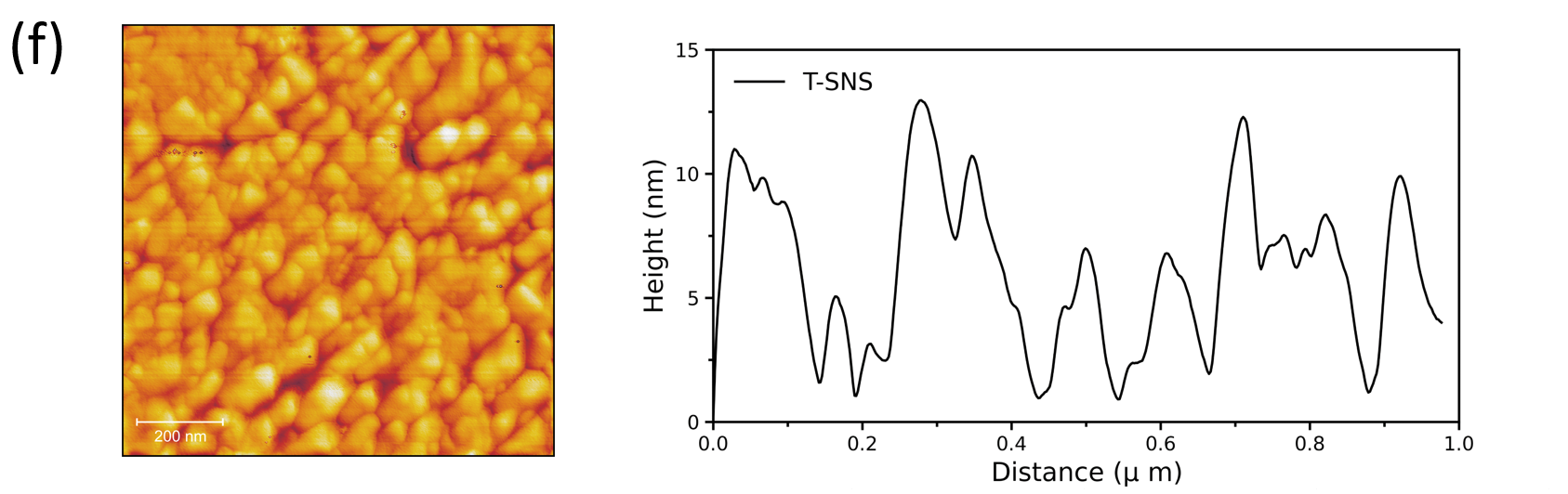}
	\end{subfigure}
	
\caption{AFM images (left) and height profiles (right) for a line drawn across the AFM images for (a) monolayer of \ce{LSMO}, (b) monolayer of \ce{LNMO}, (c) bilayer of \ce{B-NS}, (d) bilayer of \ce{B-SN}, (e) trilayer of \ce{T-NSN} and (f) trilayer of \ce{T-SNS}. The measurements were carried out using a Veeco Dimension Icon Atomic Force Microscope (AFM).}
\end{figure*}

\begin{figure*}
	\center
	\begin{subfigure}{\textwidth}
	\center
	\includegraphics[width=0.7\textwidth]{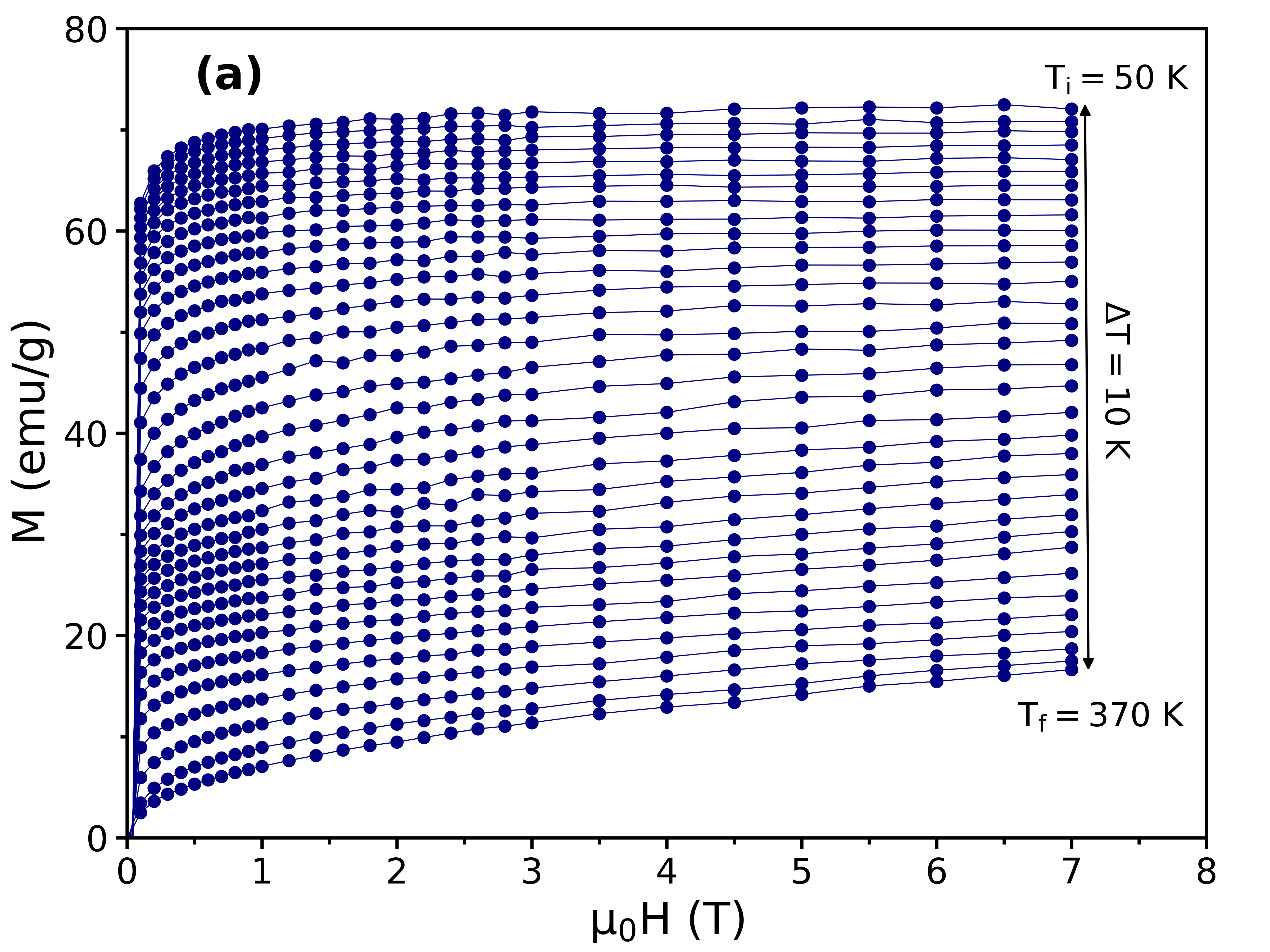}
	\end{subfigure}
\end{figure*}
\begin{figure*}
\ContinuedFloat
	\begin{subfigure}{\textwidth}
	\center
	\includegraphics[width=0.7\textwidth]{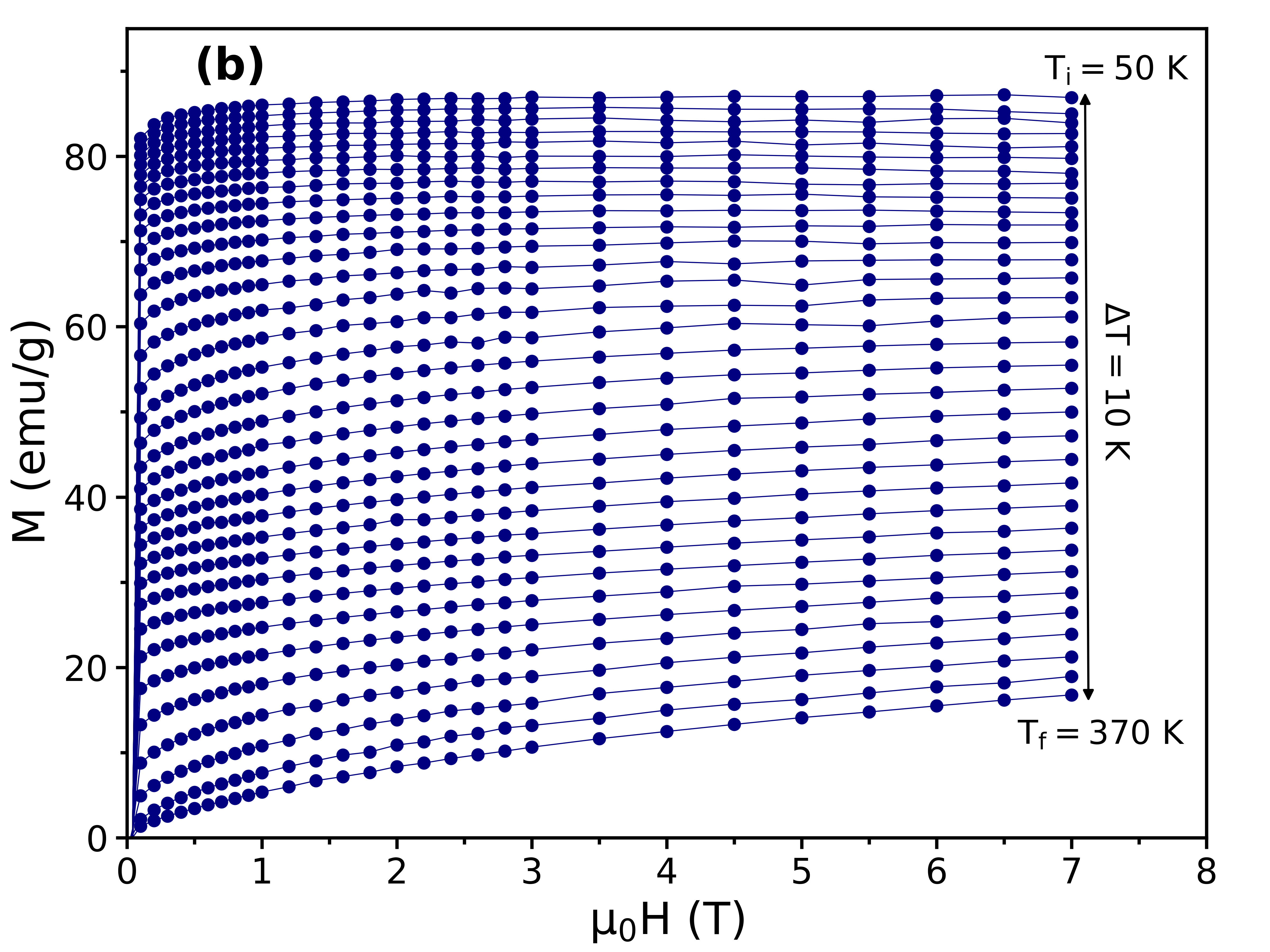}
	\end{subfigure}
\end{figure*}
\begin{figure*}
\ContinuedFloat
	\begin{subfigure}{\textwidth}
	\center
	\includegraphics[width=0.7\textwidth]{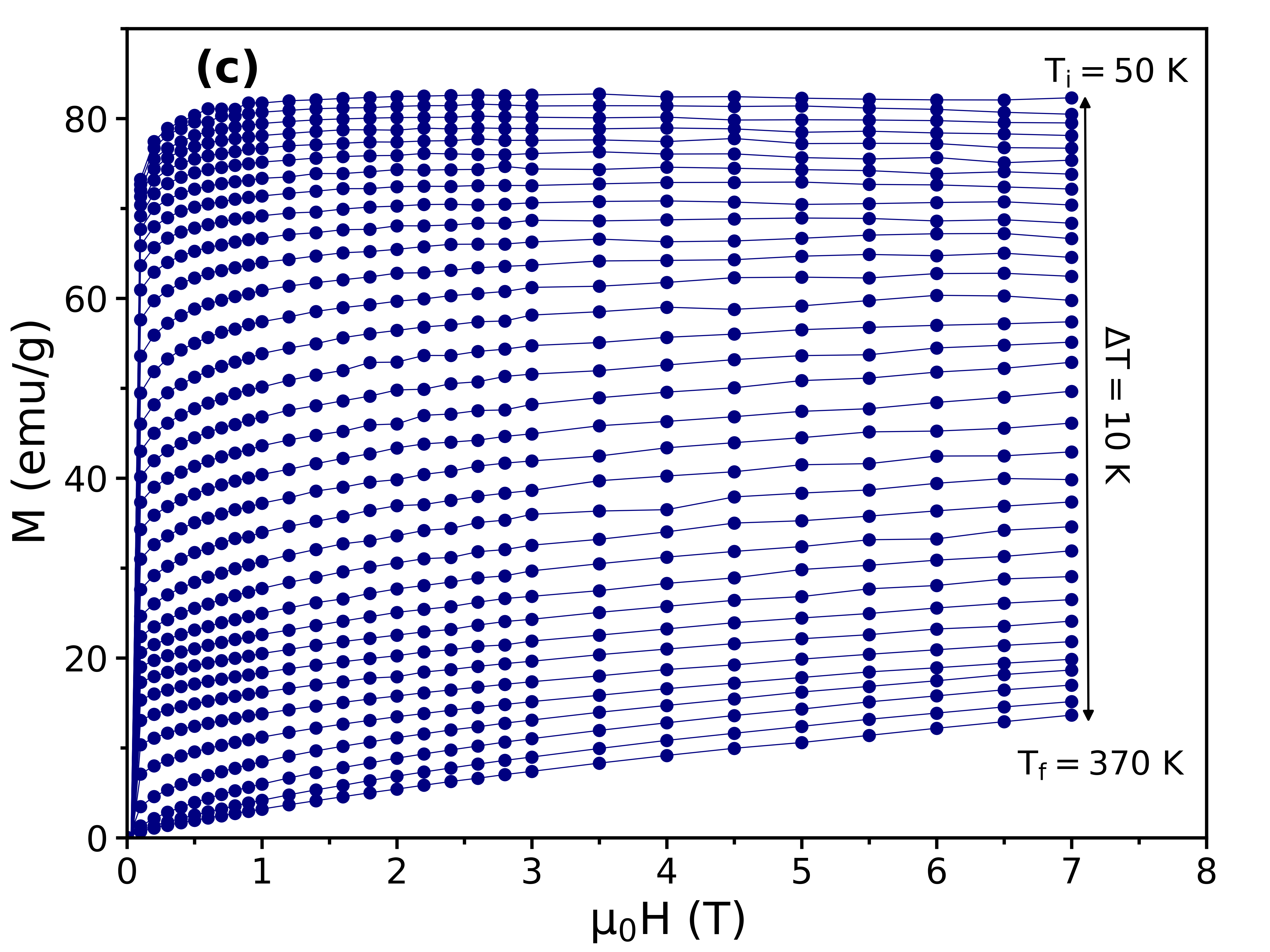}
	\end{subfigure}
\caption{Isothermal magnetization curves as a function of magnetic field in the temperature range from \SIrange[range-units=single]{50}{370}{\kelvin} with a temperature interval of \SI{10}{\kelvin} for (a) \ce{B-SN}, (b) \ce{T-SNS} and (c) \ce{T-NSN} samples. These data are obtained after the subtraction of the diamagnetic background from the substrate and the sample holder.}
\end{figure*}

\begin{figure*}
	
	\begin{subfigure}{\textwidth}
	\center
	\includegraphics[width=0.7\textwidth]{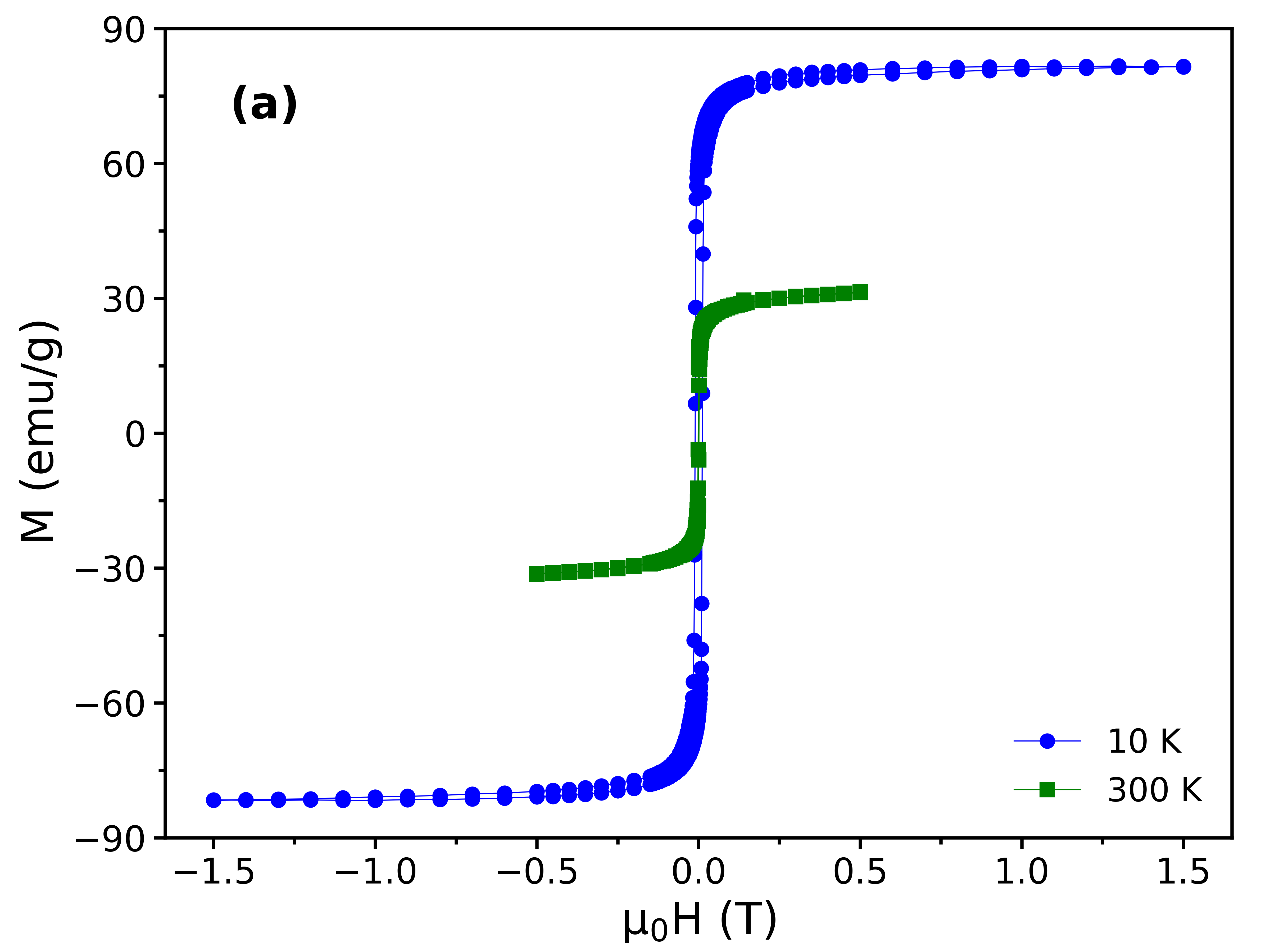}
	\end{subfigure}
\end{figure*}
\begin{figure*}
\ContinuedFloat
	\begin{subfigure}{\textwidth}
	\center
	\includegraphics[width=0.7\textwidth]{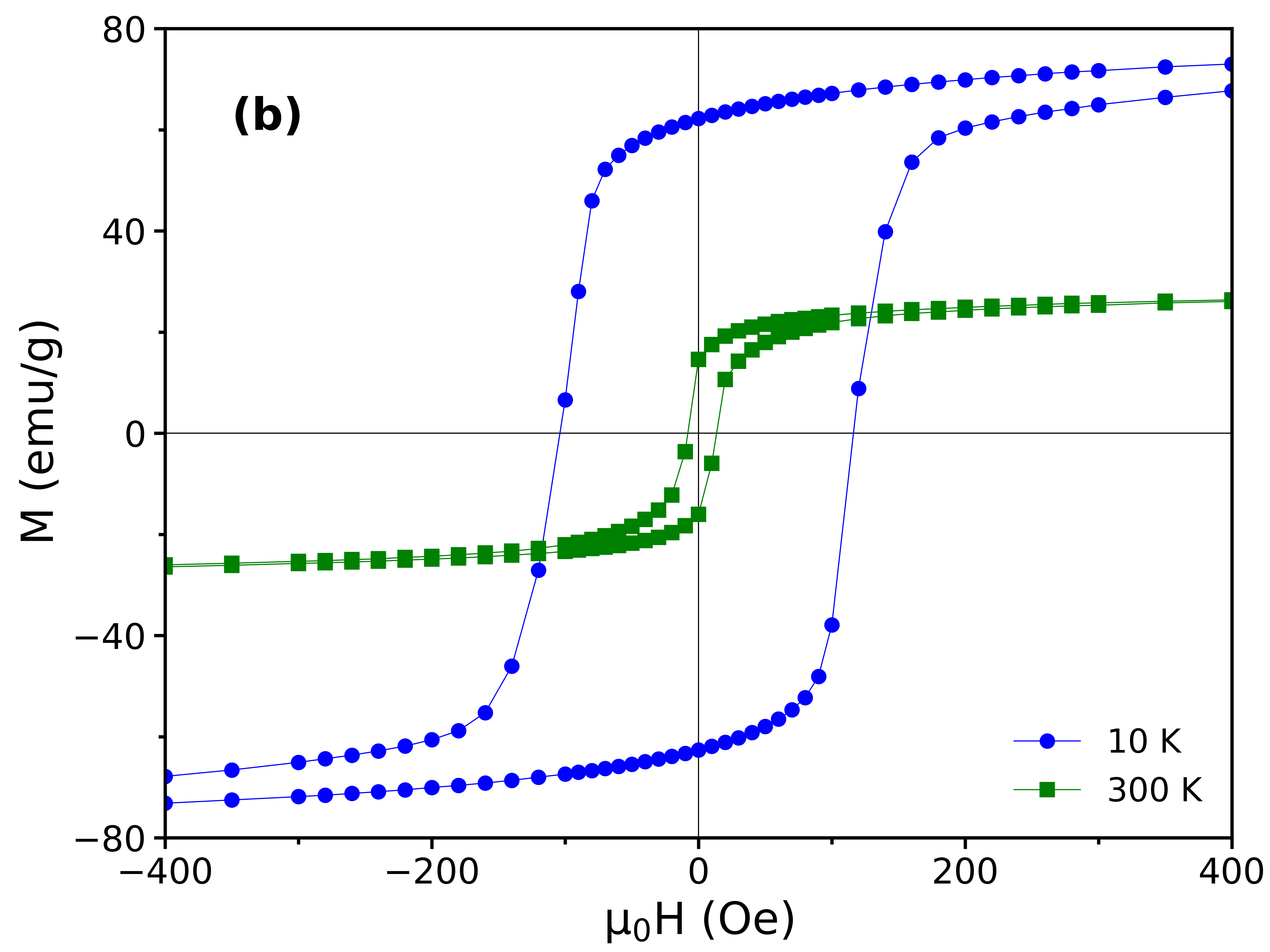}
	\end{subfigure}
\caption{(a) Magnetic hysteresis loop measured at \SIlist[list-units=single]{10;300}{\kelvin} for \ce{T-SNS} sample. (b) Magnified view around zero field showing the coercive field and the remnant magnetization at \SIlist[list-units=single]{10;300}{\kelvin}.}
\end{figure*}

\begin{figure*}
\center
\includegraphics[scale=0.6]{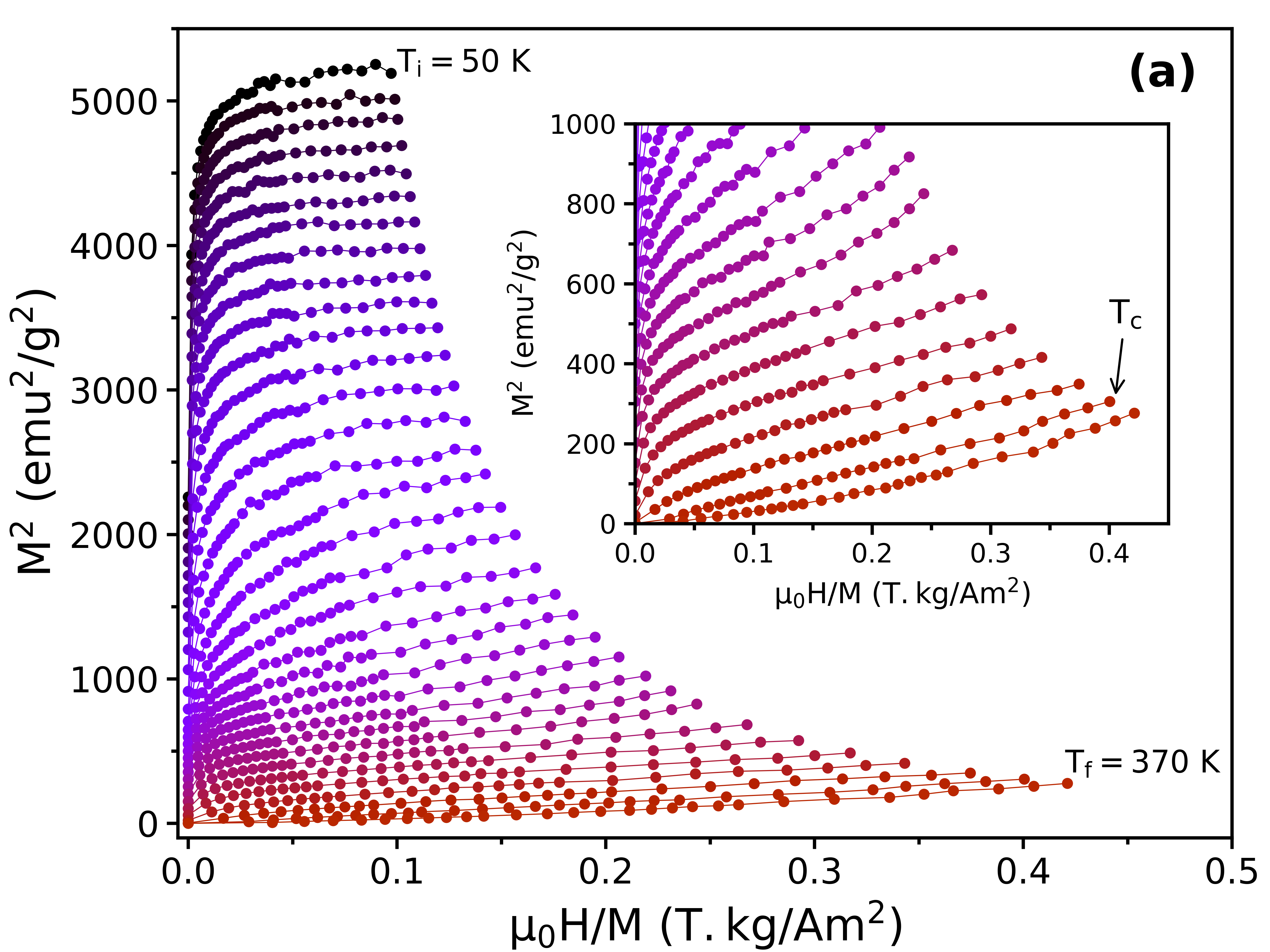}\\
\vskip25pt
\includegraphics[scale=0.6]{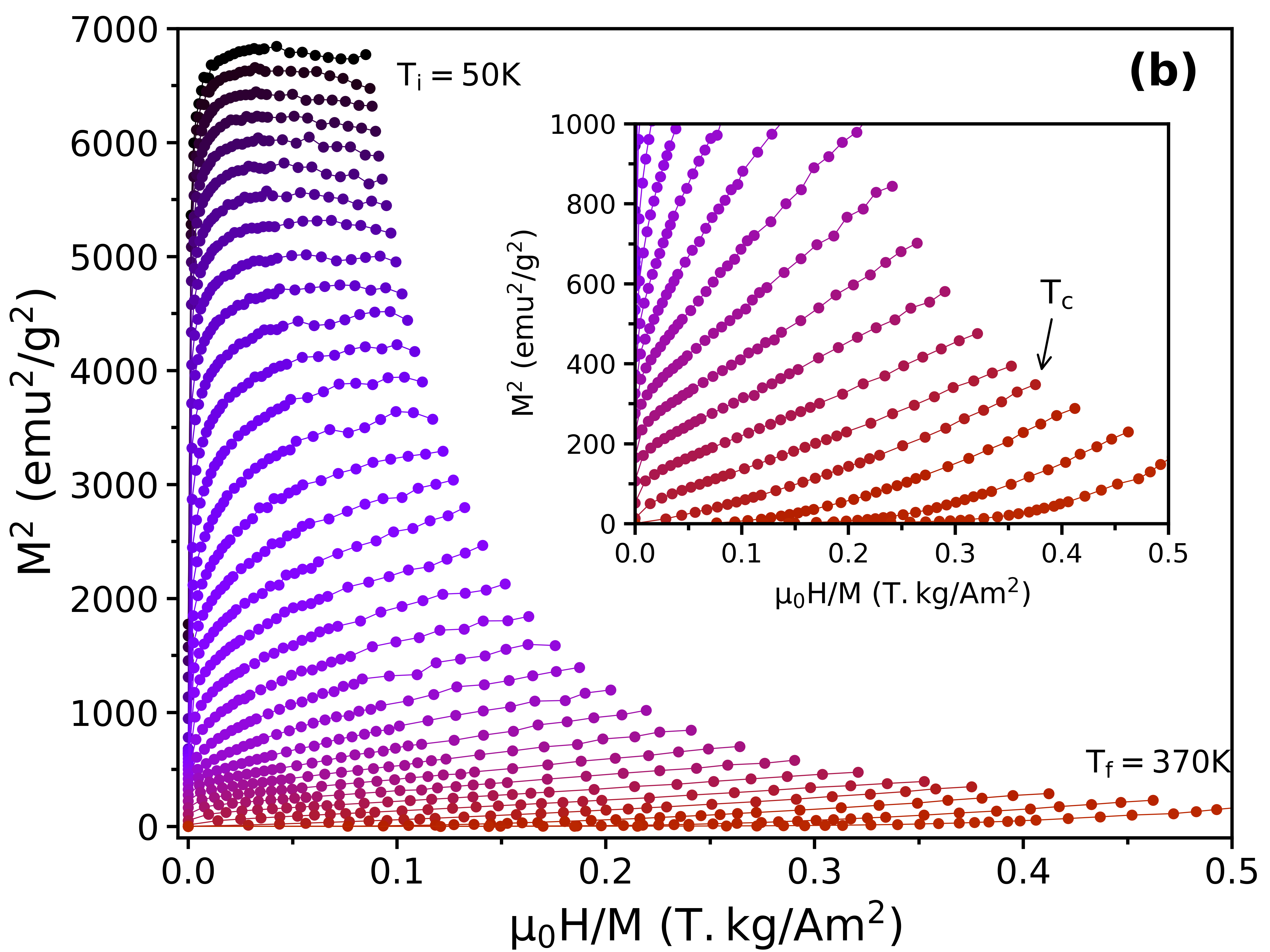}
\caption{The Arrott plots in the temperature range from \SIrange[range-units=single]{50}{370}{\kelvin} with a temperature interval of \SI{10}{\kelvin} for (a) \ce{B-SN} and (b) \ce{T-NSN} samples. Inset shows a magnified view around the transition temperature of the \ce{LSMO} layer.}
\end{figure*}

\end{document}